
\magnification = \magstep0
\nopagenumbers

\hsize=18.6truecm  \hoffset=-1.1truecm  \vsize=23.3truecm  \voffset=-0.1truecm
\hsize=18.6truecm  \hoffset=-1.1truecm  \vsize=24.3truecm  \voffset=-0.6truecm

\input ./eplain.tex

\font\mc=cmr9   \font\mit=cmti9
\font\sc=cmr8   
\font\vsc=cmr7    
\def\mathhexbox#1#2#3{\leavevmode\hbox{$\mathsurround=0pt 
                                       \mathchar"#1#2#3$}}
\def\copyright{{\ooalign
    {\hfil\raise.07ex\hbox{c}\hfil\crcr\mathhexbox20D}}}

\def\absbaselines{\baselineskip=11pt   \lineskip=0pt \lineskiplimit=0pt}
\def\sglbaselines{\baselineskip=10.4pt \lineskip=0pt \lineskiplimit=-4pt}
\def\medbaselines{\baselineskip=10pt   \lineskip=0pt \lineskiplimit=0pt}
\def\smlbaselines{\baselineskip=8pt    \lineskip=0pt \lineskiplimit=0pt}
\def\vs{\vskip 8pt} \def\vss{\vskip 6pt} \def\vsss{\vskip 2pt}
\parskip = 0pt 
\def\makeheadline{\vbox to 0pt{\vskip-30pt\line{\vbox to8.5pt{}\the
                  \headline}\vss}\nointerlineskip}

\def\footnoterule{\kern-3pt \hrule width \hsize \kern 2.6pt \vskip 3pt}

\def\omit#1{\empty}
\pretolerance=15000  \tolerance=15000
\def\ts{\thinspace}  \def\cl{\centerline}
\def\ni{\noindent}   \def\nhi{\noindent \hangindent=10pt}
       \def\bk{\kern -0.3em}  \def\b{\kern -0.1em}
\def\bks{\kern -0.05em}  \def\bbk{\kern -0.25em}  \def\bbbk{\kern -0.35em}
\def\r0{$\rho_0$}    
\def\0{\phantom{0}}  \def\00{$\phantom{000000}$}
\def\1{\phantom{1}}  
\def\etal{{\it et~al.~}} \def\huge{$\phantom{00000000000000000000000000000000}$}
\def\gapprox{$_>\atop{^\sim}$}  \def\lapprox{$_<\atop{^\sim}$}
\def\ltapprox{\hbox{$<\mkern-19mu\lower4pt\hbox{$\sim$}$}}
\def\gtapprox{\hbox{$>\mkern-19mu\lower4pt\hbox{$\sim$}$}}

\def\mltapprox{\raise2pt\hbox{$<\mkern-19mu\lower5pt\hbox{$\sim$}$}}

\newdimen\sa  \def\sd{\sa=.1em  \ifmmode $\rlap{.}$''$\kern -\sa$
                                \else \rlap{.}$''$\kern -\sa\fi}
              \def\dgd{\sa=.1em \ifmmode $\rlap{.}$^\circ$\kern -\sa$
                                \else \rlap{.}$^\circ$\kern -\sa\fi}
\newdimen\sb  \def\md{\sa=.06em \ifmmode $\rlap{.}$'$\kern -\sa$
                                \else \rlap{.}$'$\kern -\sa\fi}

\def\kms{km~s$^{-1}$}
\def\s{\ifmmode ^{\prime\prime} \else $^{\prime\prime}$ \fi}
\def\min{\ifmmode ^{\prime} \else $^{\prime}$ \fi}
\def\deg{\ifmmode ^{\circ} \else $^{\circ}$ \fi}

\def\m31{M{\ts}31}        \def\mm32{M{\ts}32}
\def\msun {M$_{\odot}$~}  \def\msund{M$_{\odot}$}  \def\mbh{$M_{\bullet}$}

\def\vphi2{v_\phi^2} \def\sphi2{\sigma_\phi^2}
\def\sz2{\sigma_z^2} \def\sr2{\sigma_r^2}

\parindent=0pt

\headline={\leftskip = -0.15in THE ASTRONOMICAL JOURNAL \hfill
           VOLUME 115, NUMBER 5 \hfill MAY 1998}

\footline={\leftskip = -0.15in \folio~~Astron.~J.~115 (5), May 1998 \hfill
           \copyright\/ 1998 Am.~Astron.~Soc.~~\folio}

\cl {\null}  

\sglbaselines

\cl{THE MASS DISTRIBUTION IN THE ELLIPTICAL GALAXY NGC 3377:} \vsss

\cl{EVIDENCE FOR A $2 \times 10^8$-\msun BLACK HOLE}

\vs

\cl{~J{\sc OHN} K{\sc ORMENDY},$^{1,\ts2,\ts3}$
    R{\sc ALF} B{\sc ENDER},$^3$
    A{\sc ARON} S.~E{\sc VANS},$^{2,\ts4}$
    {\sc AND}
    D{\sc OUGLAS} R{\sc ICHSTONE}$^5$
}

\vsss
\cl{\it Received 1997 May 30; Accepted 1998, February 4}
\vs\vs

\parindent = 37pt
\cl {ABSTRACT}
\vss\vss
{\narrower\absbaselines

\ni\quad This paper is a study of the mass distribution in the central
35\s~$\simeq$ 1.7 kpc of the E5 galaxy NGC 3377.  Stellar rotation velocity and 
velocity dispersion profiles (seeing $\sigma_* = 0\sd20$ -- 0\sd56) and $V$-band
surface photometry ($\sigma_* = 0\sd20$ -- 0\sd26) have been obtained with the
Canada-France-Hawaii Telescope.  NGC 3377 is kinematically similar to \mm32:
the central kinematic gradients are steep.  There is an unresolved central rise
in rotation velocity to $V = 110 \pm 3$ \kms~(internal error) at $r = 1\sd0$.
The apparent velocity dispersion rises from $95 \pm 2$ \kms~at $1\sd0 \leq r < 
4\s$ to $178 \pm 10$ \kms~at the center. 

\ni\quad To search for a central black hole, we derive three-dimensional 
velocity and velocity dispersion fields that fit the above observations and
{\it Hubble Space Telescope\/} surface photometry after projection and seeing 
convolution.  Isotropic models imply that the mass-to-light ratio rises by a 
factor of $\sim 4$ at $r<2\s$ to $M/L_V$ \gapprox \ts10.  If the mass-to-light
ratio of the stars, $M/L_V = 2.4 \pm 0.2$, is constant with radius, then NGC
3377 contains a central massive dark object (MDO), probably a black hole, of
mass \mbh~$\simeq (1.8 \pm 0.8) \times 10^8$ \msund.  Several arguments suggest
that NGC 3377 is likely to be nearly isotropic.  However, flattened, anisotropic
maximum entropy models can fit the present data without an MDO.  Therefore the
MDO detection in NGC 3377 is weaker than those in M{\ts}31, M{\ts}32, and NGC 
3115.

\ni\quad The above masses are corrected for the E5 shape of the galaxy and for
the difference between velocity moments and velocities given by Gaussian fits
to the line profiles.  We show that the latter correction does not affect the
strength of the MDO detection, but it slightly reduces \mbh\ and $M/L_V$.

\ni\quad At 3\s \bks\bks\bbk\lapprox $r$ \bbk\lapprox 35\s\bk, $M/L_V$ is 
constant at $\sim 2.4$. Therefore the inner parts of NGC 3377 are dominated by
a normal old stellar population.  In this elliptical, as in the bulge-dominated
galaxies NGC 3115 and NGC 4594, halo dark matter is unimportant over a 
significant range in radius near the center.


\vss
\def\huge{$\phantom{000000\ts000000000}$}


}

\parindent = 10pt

\doublecolumns\sglbaselines

\vss\vss\vskip 0.5pt
\cl {1.~\sc INTRODUCTION}
\vss

      This paper is part of a search for supermassive black holes (BHs) in 
galaxy nuclei (see Kormendy 1992a,{\ts}b, 1993; Kormendy \& Richstone 1995,
hereafter KR95; for reviews).  There is a growing body of dynamical evidence 
for central dark objects of mass \hbox{$10^{6.5}$ -- $10^{9.5}$\ts\msund} in 
galaxies.  The simplest interpretation is that these are BHs that once were 
engines for nuclear activity.  Most stellar-dynamical detections have been in 
relatively inactive bulges of disk galaxies (\m31:~Dressler 1984; Kormendy 
1987a, 1988a,{\ts}b; Dressler \& Richstone 1988; Bacon \etal 1994, NGC 
3115:~Kormendy \& Richstone 1992; Kormendy \etal 1996a, and NGC 4594:~Kormendy
1988c; Emsellem \etal 1994; Kormendy {\it et al.}~1996b).  In contrast, nuclear
activity is strongest in giant ellipticals.  The only ellipticals with 
stellar-dynamical evidence for BHs are the inactive dwarfs \mm32 (Tonry 1984, 
1987; Dressler \& Richstone 1988; van der Marel \etal 1994b; Qian \etal 1995;
Dehnen 1995; Bender, Kormendy, \& Dehnen 1996; van der Marel \etal 1997a,{\ts}b)
and NGC 4486B (Kormendy \etal 1997).  However, failure to detect BHs is not 
evidence against them.  Gas-dynamical searches have found BHs in six
\huge\vskip -10.4pt

\vss\vss\vskip -1.5pt

\hrule width 4.2truecm

\vss\vskip -1pt

{\vsc\smlbaselines

$^1${\ts}Visiting Astronomer, Canada--France--Hawaii Telescope, operated by the
    National Research Council of Canada, the Centre National de la Recherche
    Scientifique of France, and the University of Hawaii.

$^2${\ts}Institute for Astronomy, University of Hawaii, 2680 Woodlawn Dr., 
    Honolulu, HI 96822; Electronic mail: kormendy@ifa.hawaii.edu

$^3${\ts}Universit\"ats-Sternwarte,{\ts}Scheinerstra\ss e{\ts}1,{\ts}M\"
    unchen{\ts}81679,{\ts}Germany; Electronic mail: bender@usm.uni-muenchen.de

$^4${\ts}Present address: Caltech 105-24, Pasadena, CA 91125; Electronic mail:
      ase@astro.caltech.edu

$^5${\ts}Department of Astronomy, University of Michigan, Ann Arbor, MI 48109;
     Electronic mail: dor@astro.lsa.umich.edu

}

\ni  galaxies, three of which are active giant ellipticals (Harms \etal 1994;
Ferrarese \etal 1996; Bower \etal 1998).  In fact, any giant
elliptical could hide a $10^8$-\msun BH from past stellar-dynamical searches.

      The BH search is becoming dominated by the {\it Hubble Space Telescope\/}
(\b {\it HST\/}); with it, BH detection is possible in most giant galaxies out
to the distance of the \hbox{Virgo cluster.}  But for ground-based searches, BH
detection is difficult.  There are two main reasons.  First, giant ellipticals
do not rotate, so mass measurements are maximally sensitive to velocity 
anisotropies.  Second (e.{\ts}g., Lauer \etal 1995), they have cuspy cores with
large break radii; inside $r_b$ the brightness profile is relatively shallow, so
projected spectra are dominated by light from large radii where a BH has no 
effect.  Kormendy (1992b) illustrates this effect for the {\it HST\/} profile of
M{\ts}87.  

      The BH in \mm32 was found because the galaxy is nearby and rapidly
rotating.  Similarly, \m31, NGC 3115, and NGC 4594 contain rapidly rotating 
nuclear disks.  They are nearly edge-on; this ensures that spatially unresolved
rotation contributes to the apparent velocity dispersion.  Finally, a detection 
was possible in NGC 4486B because the BH is unusually massive compared to the
rest of the galaxy.  Ground-based BH detection is still only possible in
galaxies like these where circumstances are favorable.  

      NGC 3377 is such a galaxy.  It is a prototypical E5 galaxy illustrated 
in the {\it Hubble Atlas} (Sandage 1961).  At $M_B = -18.8$, it is just fainter
than the transition between ellipticals that rotate and those that do not: it 
rotates rapidly enough to be nearly isotropic (Davies \etal 1983).  Also, it
has a coreless, power-law profile (Lauer \etal 1995), so small radii have large
luminosity weight in projection.  The axial ratio is 0.5; since no elliptical 
is much flatter (Sandage, Freeman, \& Stokes 1970; Binney \& de Vaucouleurs
1981; Franx, Illingworth, \& de Zeeuw 1991; Tremblay \& Merritt 1995), NGC 3377
must have a high inclination.  We assume that it is edge-on.  Finally, NGC 3377
is one of the nearest ellipticals. It is therefore an excellent target for a BH
search.  This paper presents high-resolution surface photometry and 
stellar-kinematic measurements and uses them to calculate the mass-to-light 
ratio $M/L_V$ as a function of radius.  NGC 3377 turns out to be kinematically 
similar to \mm32.  Isotropic models imply that it contains a BH of mass 
\mbh~$\simeq 1.8$ $\times \ts10^8$ \msund.

     We also discuss the mass distribution at 3$^{\prime\prime}$\bks\ts \lapprox
\ts$r$ \lapprox \ts35$^{\prime\prime}$.  Mass distributions have been measured 
in only a few bulges and ellipticals.  At radii well outside the de Vaucouleurs 
(1948) effective radius $r_e$, a variety of techniques show that halo dark 
matter dominates the mass distribution (see Kent 1990 for a review).  
Mass-to-light ratios $M/L_V \sim 10^2$ are large compared to values $M/L_V$ 
\lapprox \ts10 for old stellar populations.  Measurements at $r$ \lapprox 
\ts$r_e$ need to be more accurate.  Variations in $M/L_V$ due to metallicity
gradients, halo dark matter and all but the most spectacular BHs are likely to 
be smaller than a factor of 2 to 4.  Measurements intended to be this precise 
are vulnerable to velocity anisotropies.  Only a few well-studied galaxies are
likely to be isotropic or else contain embedded, edge-on disks that simplify the
measurements without contributing much mass (e.{\ts}g., NGC 3115 and NGC 4594,
see Kormendy \& Westpfahl 1989 for the latter).  The observed result is that 
these galaxies have $M/L_V \simeq$ constant near their centers.  Moreover, 
$M/L_V$ is small, so the mass is dominated by an old stellar population and not
by halo dark matter.  Here we show that this is also the case in NGC 3377.

\headline={\vbox to 0pt{\leftskip = -0.15in 
        \folio~~~KORMENDY {\it ET AL.}: BLACK HOLE IN NGC 3377 \hfill \folio}}

\footline={}

  \omit{This paper gives only a brief description of the BH search technique.
Each step is discussed in detail in previous papers, especially Kormendy and 
Richstone (1991). Section 2 discusses the kinematic measurements and takes a 
first look at the results.  Section 3 presents the surface photometry.  Given 
measurements of projected quantities, the most important part of the analysis 
then is the derivation of unprojected brightnesses (\S{\ts}5) and rotation 
velocities and velocity dispersions (\S{\ts}6 for isotropic models; \S{\ts}7 for
anisotropic models).  Section 8 discusses other degrees of freedom on the mass 
distribution and sums up.
      This paper is really two papers in one.  Readers interested in the 
structure of NGC 3377 as a typical low-luminosity elliptical can concentrate on
\hbox{\S\S\ts2\ts--\ts4, 6, 9, and the Appendix.}  NGC 3377 as a black hole 
candidate is discussed in \S\S\ts2 and 5 -- 9.}

      We assume that the distance to NGC{\ts}3377 is 9.9{\ts}Mpc based on a 
Hubble constant of $H_0 = 80$ km s$^{-1}$ Mpc$^{-1}$, the recession velocity 
$626 \pm 39$ \kms~of the Leo Spur (Group 15 $-1$ in Tully 1988) and a  
Virgocentric flow solution (Faber \etal 1997).  Then the scale is 
20\sd8 kpc$^{-1}$.  The Galactic absorption is assumed to be $A_B = 0.06$ 
(Burstein \& Heiles 1984).

\vskip 9.5truecm

\includegraphics{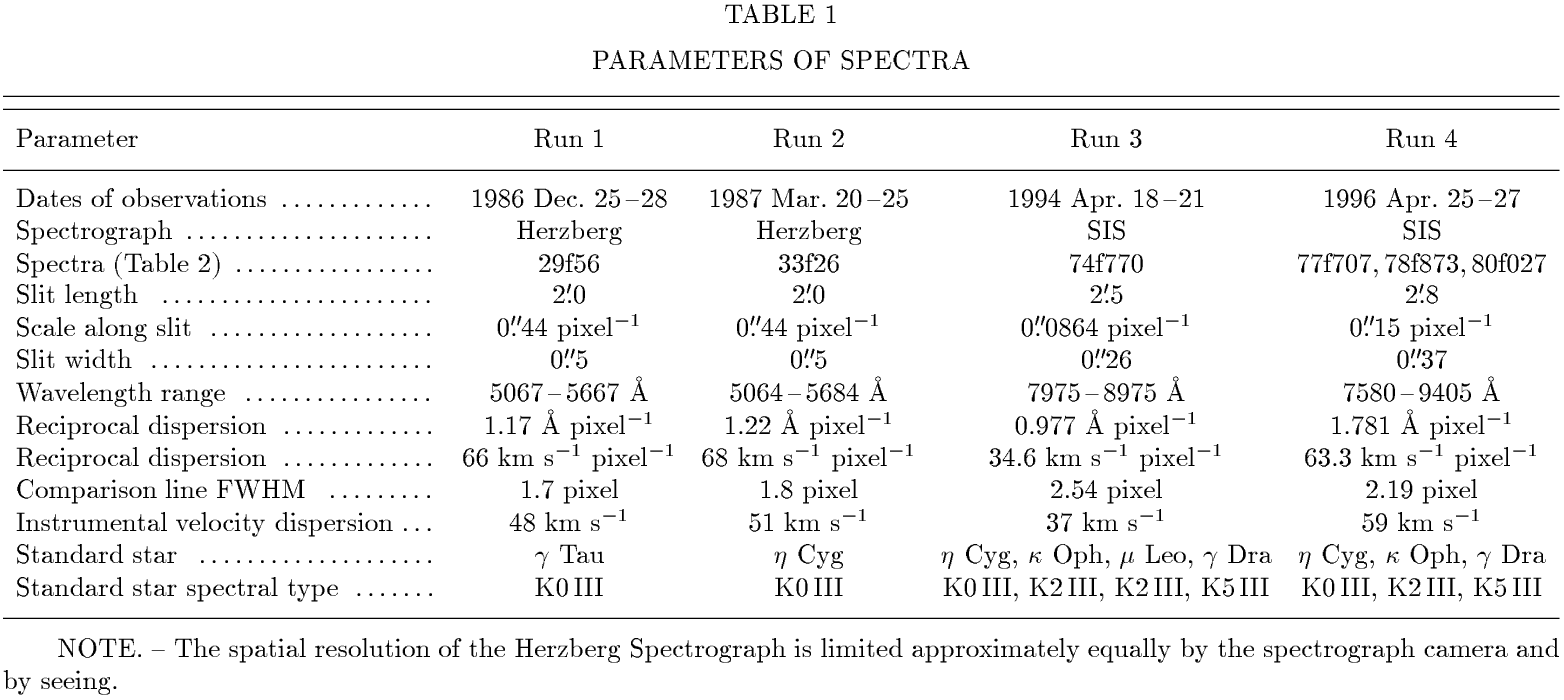}

\vss\vsss\vsss\vskip -2pt
\cl {2.~\sc KINEMATICS}

\vss
\cl {2.1.~\it Observations and Data Reduction}
\vss

      The kinematic measurements were made during four observing runs with the
Canada-France-Hawaii Telescope.  For the first two runs, the Herzberg 
Spectrograph (Salmon 1985) was used at f/4 with an RCA CCD (316 $\times$ 498, 
30 $\mu$m pixels; read noise \hbox{$\sim 71$ e$^{-1}$ {\bk}pixel$^{-1}$;} Walker
\etal 1984).  Spectra were taken at position angles PA = 50$^{\circ}$ and
44$^{\circ}$.  Leach (1981) measured these as the PAs of the major axis at 
small and moderate radii.  Subsequently, we found that our photometry shows no
twist near the center.  Present and published CCD photometry (\S{\ts}3) imply 
that the major axis is at PA $= 41^{\circ} \pm 1^{\circ}$ (external error).  We 
neglect the small difference between the correct major-axis PA and that of the 
spectra.  

      For subsequent runs, we used the Subarcsecond Imaging 
Spectrograph (SIS).  Tip-tilt guiding is incorporated; by offsetting the guide
probe, we can center the object on the slit to one-pixel accuracy.  An observing
sequence consists of a series of direct images to center the object at the slit,
an exposure with the slit in place but with no grism, the object spectrum, 
another image through the slit but without the grism and one with neither slit 
nor grism to verify that the object is still centered, and a comparison 
spectrum.  The seeing was measured on the bracketing direct images.  The 
brightness profile of the galaxy is the same in these images and in the 
spectrum, so the PSF in the images is correct for the spectrum.

      Parameters of the spectra are given in Table 1.  Integration times, 
position angles, and seeing estimates are given in Table 2.

\omit{
      The spatial resolution was measured by taking a 180 -- 300 s guided
integration on a ``seeing star'' near NGC 3377 immediately after each galaxy 
exposure.  NGC 3377 was observed during the same runs as NGC 3115; Kormendy \&
Richstone (1992) found that the ratio of FWHM values for seeing and focus stars
is FWHM$_{\rm seeing}$/FWHM$_{\rm focus} = 1.06 \pm 0.03$.  This implies that 
guiding errors do not greatly affect the resolution. Guiding on NGC 3377 is easy
because the core is small and bright.  The measured $\sigma_* =$ FWHM$_{\rm 
seeing}$/2.35 is given along with the kinematic results in Table 2.  The actual
seeing was better; the spatial resolution is limited by the spectrograph. }

      Atmospheric dispersion was negligible.  The centering exposures were 
taken with an $I$ filter and the same CCD that was used for the spectroscopy, 
so the effective wavelength was close to that of the Ca triplet. Also, the mean
zenith distances were 24\dgd3 for spectrum 77f707, 6\dgd0 for spectrum 78f873, 
and 7\dgd5 for spectrum 80f027.  The BH models were fitted to spectrum 80f027.

\vfill\eject


      Instrumental reduction of RCA CCD data is routine.  The spectra were 
corrected for geometric distortion and rewritten on a $\ln {\lambda}$ scale 
using the ``longslit'' package in the National Optical Astronomy Observatories'
{\it Image Reduction and Analysis Facility} (Tody \etal 1986). 

      Spectra of MK standard stars chosen from Morgan \& Keenan (1973) were
observed and reduced similarly.  Since their images were smaller than the slit,
stars were trailed along the slit at position angles slightly different from 
0$^{\circ}$ or 90$^{\circ}$.  After rectification, intensities were averaged 
along the slit to produce one-dimensional spectra with the proper instrumental 
dispersion.  In the Fourier analysis, different K0{\ts}--{\ts}5 III stars gave 
almost identical results.  The adopted star(s) (Table 1) gave marginally the
best internal and external consistency.

      Velocities $V$ and velocity dispersions $\sigma$ were calculated using a 
Fourier quotient program (Sargent \etal 1977; Schechter \& Gunn 1979) as
discussed in Kormendy \& Illingworth (1982) and in Kormendy \& Richstone (1992).
Results are listed in Table 2.  The adopted center is determined with an 
accuracy of $\sim 0.1$ pixel by comparing the brightness profile along the slit
with the surface photometry discussed in \S{\ts}3.  The rotation curves measured
from the Run 4 spectra are slightly asymmetric; the results in Table 2 contain a
shift of $\Delta r = 0\sd15$ applied to all radii.  It is not clear whether this
is real (as in M{\ts}31) or not.  It is difficult to believe that it could be 
due to an undiagnosed problem with centering or guiding, because the brightness
profile along the slit provides a check on both.  We will not attempt to 
interpret the asymmetry; it is small enough that it does not affect our 
conclusions.

\vss\vsss\vsss\vsss\vsss\vsss\vsss\vskip 1pt
\cl {2.2.~\it A First Look at the Kinematics}
\vs\vsss\vsss

     Figure 1 illustrates the kinematics.  The rotation and dispersion profiles
closely resemble those of \mm32~(Tonry 1984, 1987; Kormendy 1987a; Dressler \& 
Richstone 1988; Carter \& Jenkins 1993; van der Marel \etal 1994a; Bender, 
Kormendy, \& Dehnen 1996; van der Marel, de Zeeuw, \& Rix 1997).  The kinematic
gradients near the center are unresolved.  The maximum rotation velocity $V = 
110 \pm 3$ \kms~(internal error) has already been reached 1\sd0 from the center.
The velocity dispersion increases by 87\ts\% from $95 \pm 2$ \kms~at $1\sd0 \leq
r<4\s$ to $178 \pm 10$ \kms~at the center.  As in M{\ts}32, $V$ and $\sigma$ 
continue to decrease slowly at large radii.

      Our value of the central dispersion is somewhat larger than $\sigma = 160$
\kms~adopted by Whitmore, McElroy, \& Tonry (1985) based on three published 
measurements.  We also find a larger maximum rotation velocity than the value
$80 \pm 6$ \kms~quoted by Davies \etal (1983) based on measurements in
Illingworth (1977).  The difference is due to the lower resolution of the
photographic spectra available in 1977.  Figure 1 shows that the maximum 
rotation velocity, the apparent central velocity dispersion, and the central 
gradients in $V$ and $\sigma$ all get larger as resolution improves. 

\huge

\singlecolumn

\vskip 12.5truecm

\includegraphics{./3377fig1.cps}

{\medbaselines\mc{F{\sc IG}.~1.}---Folded rotation curve and dispersion profile
along the major axis of NGC 3377.  Central points of SIS spectrum 80f027 are
separated by 0\sd15.  }

\vfill\eject

\doublecolumns

\cl\null

\includegraphics{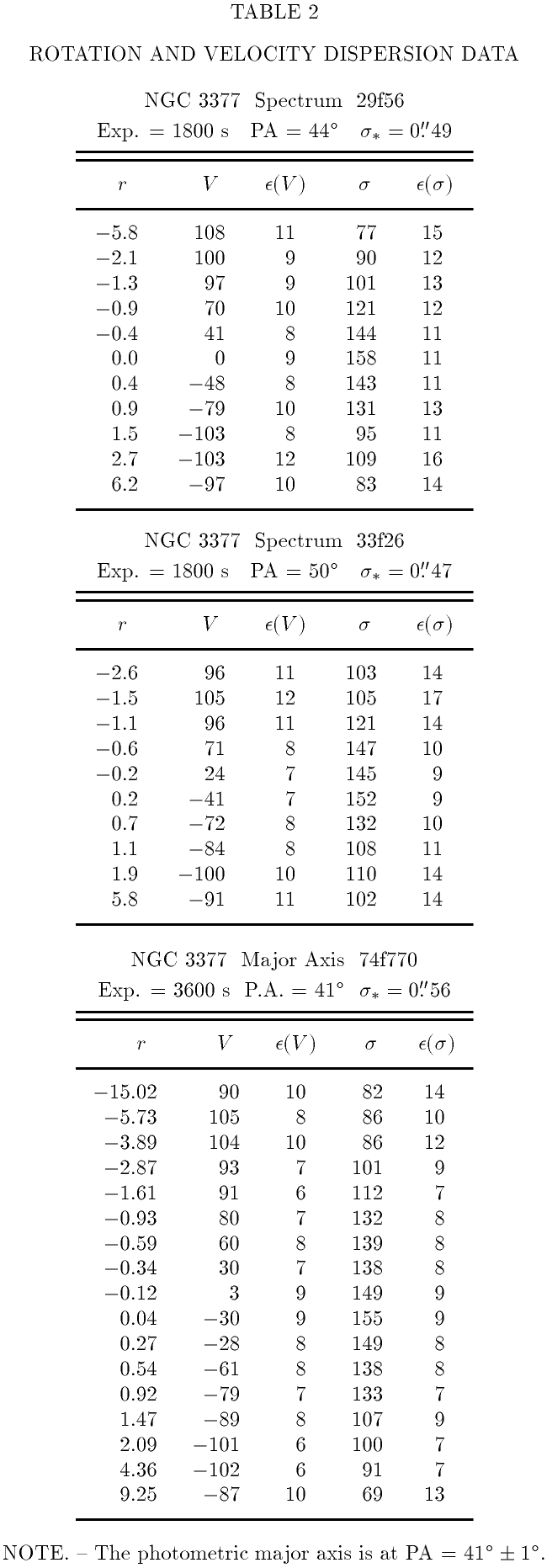}
\includegraphics{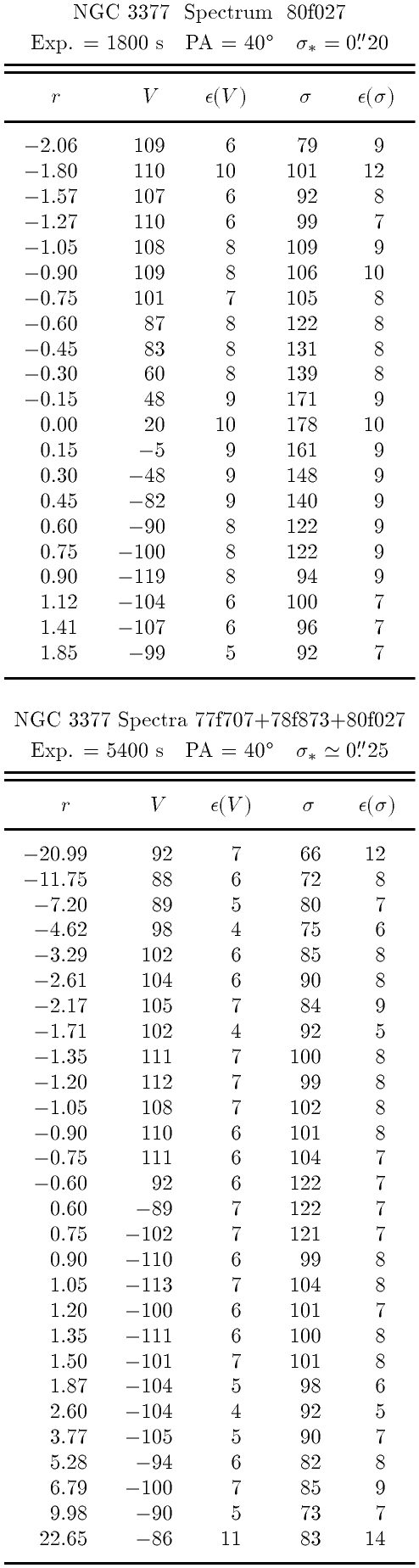}

\vfill\eject

\singlecolumn
\cl{\null}

\vfill

\includegraphics{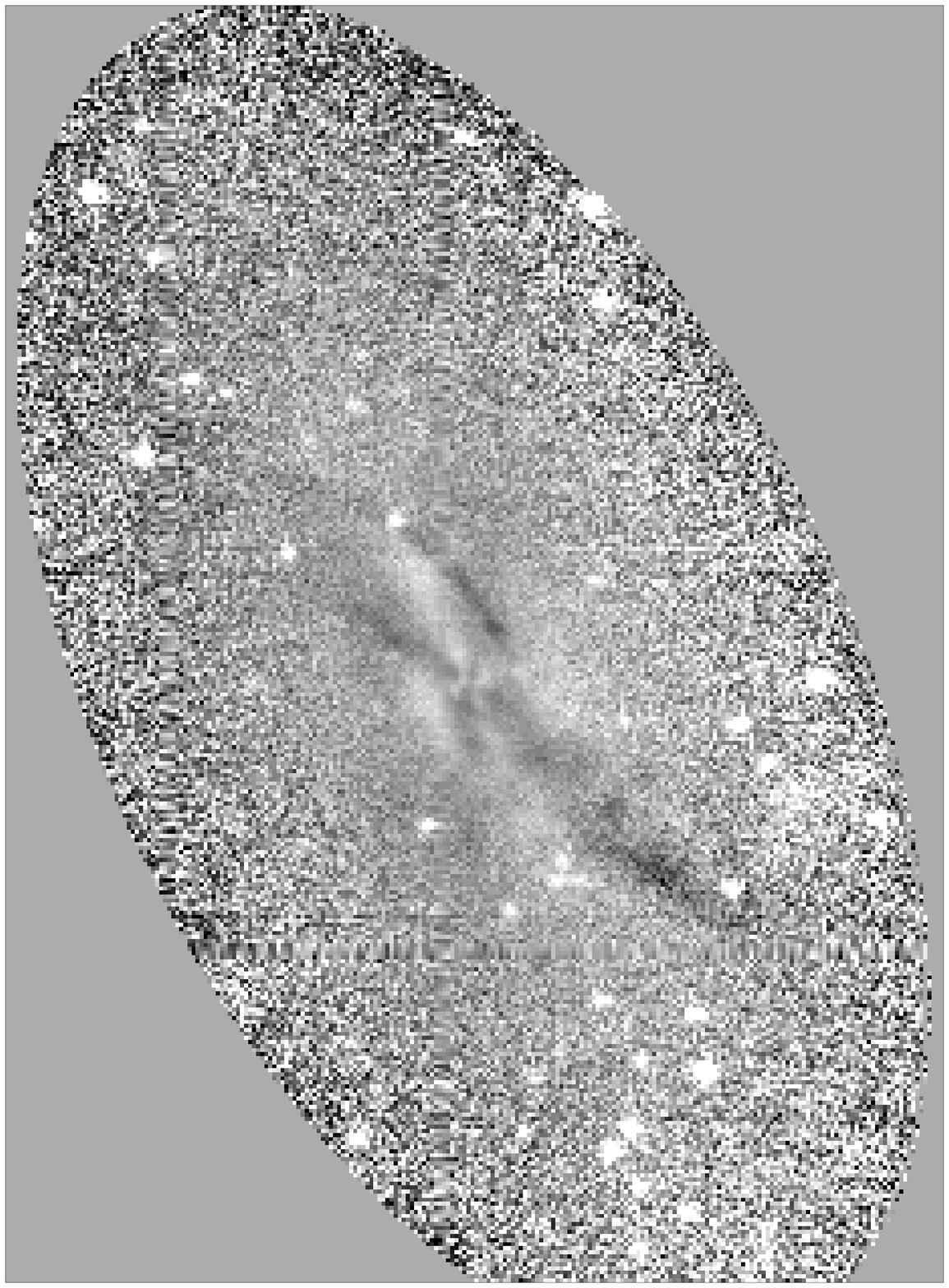}

{\medbaselines\mc{F{\sc IG}.~2.}---Unsharp-masked, $V$-band image  of NGC 3377.
This is image 5f57 in Table 3, i.{\ts}e., a 180 s exposure with seeing 
$\sigma_* = 0\sd26$.  Inside the outermost contour for which the brightness
profile was calculated, the image was divided by a reconstruction that has
exactly elliptical isophotes and the galaxy's measured brightness, ellipticity,
and PA profiles.  Outside this contour, the image is set equal to 1.  The
grayscale is linear between 0.8 (black) and 1.1 (white).  Most dust patches are
a few percent deep; the strongest absorption is 8\ts\%.  Note that the
appearance of diskiness is enhanced by the geometry of the dust distribution.
The area shown is 1\md13 high and 0\md83 wide.  North is 16$^\circ$ clockwise
from upward, and east is 16$^\circ$ clockwise from left.}

\eject

\doublecolumns

\vss\vsss\vsss
\cl {3.~\sc PHOTOMETRY}
\vsss\vsss\vskip 1pt

      Four $V$-band images of NGC 3377 were obtained at the CFHT Cassegrain
focus with the RCA CCD (scale $=$ 0\sd215 pixel$^{-1}$). Results
are listed in Table 3.   For images 5f54 --
5f57, the quoted seeing is an average for 0, 1, 3, and 9 stars, respectively,
scattered around the galaxy on the frames.  No star is bright enough for the
shortest exposure, 5f54.  However, Figure 4 (below) clearly shows that the 
seeing was better than for 5f55, which has a well measured $\sigma_* = 0\sd22$.
Also, at the time of the observations (1984 March), the best images ever 
obtained with the CFHT had $\sigma_* \simeq 0\sd19$ (FWHM $\simeq$ 0\sd45).
Therefore we adopt $\sigma_* = 0\sd20$ $\pm$ 0\sd01.

      CCD photometry of large galaxies is usually limited by the accuracy of sky
subtraction.  Here, too, the galaxy image is larger than the chip.  For all four
images, we adopt the median sky brightness in a 900 s exposure on a blank field
taken immediately after image 5f57.  We also measured  median brightnesses at 
the corners of the galaxy exposures (in each case, NGC 3377 is close to the 
center).  The ratios of these brightnesses to the adopted sky values are 1.18, 
$1.17 \pm 0.06$, $1.16 \pm 0.01$, and $1.17 \pm 0.01$ for 5f54 -- 5f57, 
respectively.  This consistency implies that the sky brightness did not vary 
significantly during the 35 m course of the galaxy and sky exposures.  For 
images 5f54 -- 5f57, Table 3 lists profiles out to only 2.7 -- 1.1 times the sky
brightness.  We have not tried to measure NGC 3377 to faint levels because the 
sky brightness is less certain than if it were measured on the object images.
We will in any case use a composite profile that at large radii is based on 
photometry with large-field CCDs.  Also, our analysis is insensitive to the 
outer profile.  

      The profiles were calculated using a slightly modified form of the PROFILE
program in the Lick Observatory image processing system VISTA (Stover 1988). 
PROFILE was written by T.~Lauer (1985).  It uses sinc interpolation optimized
for high spatial resolution and is remarkably accurate.  Even test profiles 
observed with 2 -- 3 pixels per FWHM are well measured.  

      PROFILE fits elliptical isophotes to the image.  Since real isophotes in
many elliptical galaxies show disky or boxy distortions, we checked that the 
above procedure adequately measures the major-axis profile.  Figure 2 shows an 
``unsharp-masked'' version of image 5f57.  A synthetic image was constructed 
with exactly elliptical isophotes and the measured brightness, ellipticity, and
PA profiles. Image 5f57 was divided by this synthetic image to produce Figure 2.
The cross-shaped pattern is the signature of a disky distortion (Carter 1987; 
Jedrzejewski 1987; Michard \& Simien 1988; Bender, D\"obereiner, \& M\"ollenhoff
1988; Nieto \& Bender 1989; Peletier \etal 1990; Nieto \etal 1991; Scorza \& 
Bender 1995); this is normal for an elliptical galaxy that rotates rapidly
(Bender 1988; Nieto, Capaccioli, \& Held 1988; Nieto \& Bender 1989; Bender 
\etal 1989; Kormendy \& Bender 1996).  Most of the light in the disky distortion
is included in the profile; the residuals are too small to affect our analysis.
Figure 2 also shows an irregular distribution of dust, but the amount of 
absorption is too small to be important here.

      The zeropoint of the $V$ magnitude scale \hbox{is based on} published 
aperture photometry by Webb (1964); Strom \etal (1976); Sandage \& Visvanathan 
(1978); Persson, Frogel, \& Aaronson (1979); Caldwell (1983), and Poulain 
(1988).~Ten measurements with aperture radii of 11\sd5~to 30\sd5 were used.  One
measurement by Webb through a 12\sd3 \omit{radius} aperture was discarded; it
appears imperfectly centered.  For each measurement, synthetic aperture 
photometry on image 5f57 gave an instrumental magnitude.  The difference between
this and the published magnitude is the zeropoint.  Zeropoints were found to be
independent of aperture radius.  Results from different papers are remarkably 
consistent: the mean of ten zeropoint determinations has an accuracy of 
$\pm 0.004$ mag.  We applied the zeropoint from image 5f57 to the other images 
by shifting the profiles in $\mu$ to minimize the scatter.  We also 
checked that zeropoints calculated from the individual images are consistent.
For images 5f54 -- 5f56, they differ from the adopted zeropoint by 0.002, 0.025,
and 0.002 mag arcsec$^{-2}$, respectively. Table 3 lists the brightness profiles
with 5f57 zeropoints applied.  Major-axis radius $r$ is in arcsec, $\mu_V$
is in $V$ mag arcsec$^{-2}$, $\epsilon$ is ellipticity, and PA is position angle
in degrees east of north.

      We verified that, except for resolution differences, the galaxy brightness
profile is the same on the spectra and on the images. 

    Figure 3 shows the results. Also shown is photometry by Jedrzejewski (1987),
Michard \& Simien (1988), Peletier \etal (1990), and Pierce (1991).  {\it HST\/}
WFPC1 photometry from Lauer \etal (1995) provides the profile near the center.
Figure 3 is the most accurate composite profile we can derive from available 
data; the published profiles are truncated at small radii where they ``peel 
off'' because of low resolution.  The profiles by Jedrzejewski (1987) and by 
Peletier \etal (1990) are also truncated at large radii because of deviations 
due to inaccurate sky subtraction. The best data at large radii are those by
Pierce and by Michard and Simien.  Pierce used an 800 $\times$ 800 pixel TI CCD
at the University of Hawaii 2.2 m telescope; the scale was 0\sd6 pixel$^{-1}$
and the unvignetted field was 5\min \bk$\times$ 5\min\bbbk.  The sky was 
determined at the corners of the images; sky subtraction should be more accurate
than for the other CCD photometry.  Michard and Simien took photographic plates 
at the 1.2 m telescope of the Observatoire de Haute-Provence.  The CCD data are
preferred at small radii, but the photographic profile is probably the most 
accurate at large radii.

\vfill

\includegraphics{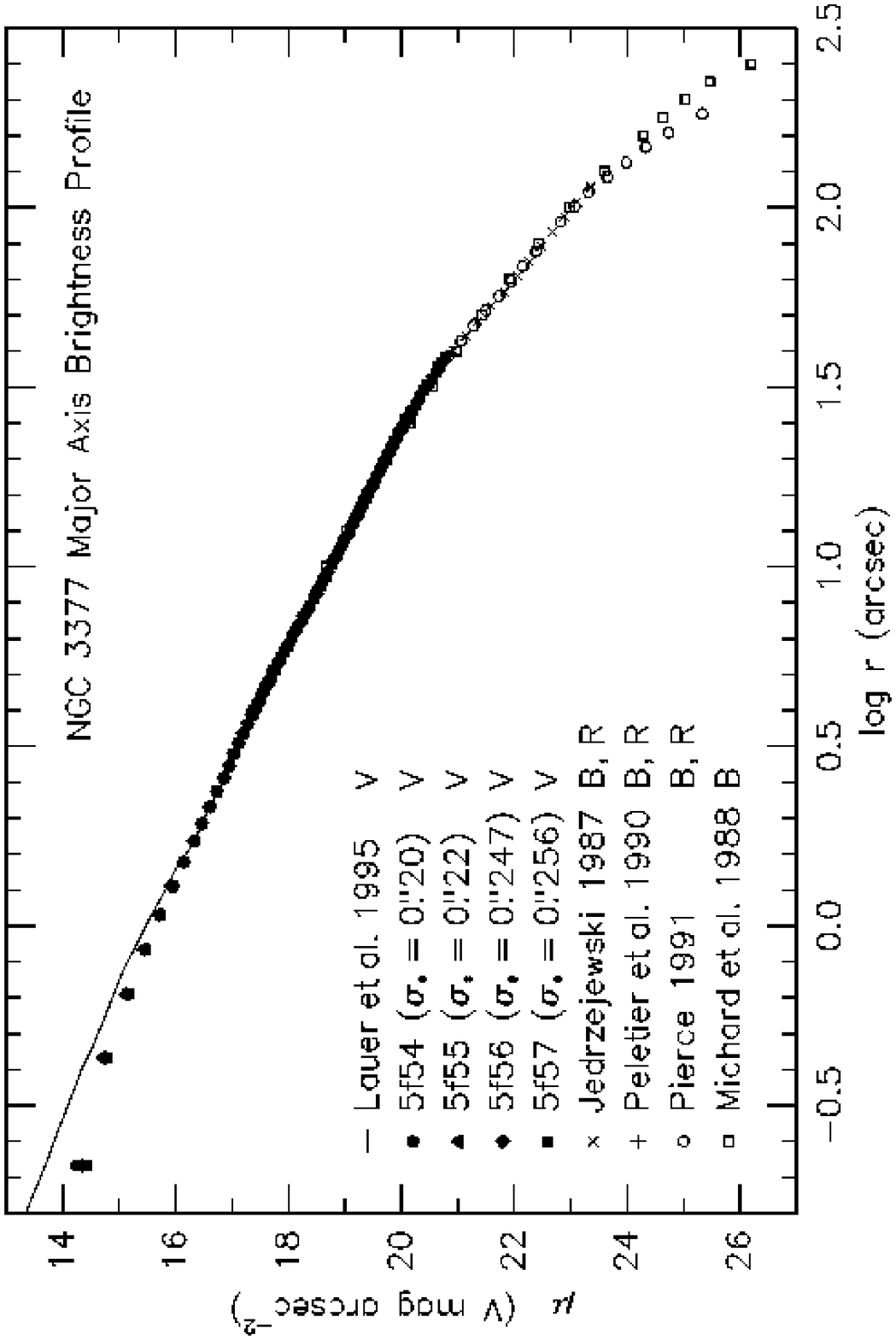}

{\medbaselines\mc{F{\sc IG}.~3.}---Major-axis brightness profile of NGC 3377. 
Sources of the published photometry are given in the key; 
``$B$,$R$\ts'' means that $B$- and $R$-band profiles have been 
intensity-averaged to approximate $V$.  Since points are not distinguishable at 
$\log {r} < 1.6$, we note that the log radius range of the plotted data is: for
Jedrzejewski (1987), 0.81 -- 2.06; for Peletier \etal (1990), 0.51 -- 1.40; for
Pierce (1991), 0.85 -- 2.26; and for Michard \& Simien (1988), 1.00 -- 2.40.
Points omitted because of poor seeing or sky subtraction are included in Figure
4.}

\eject

\singlecolumn

\cl{\null} \vfill

\includegraphics{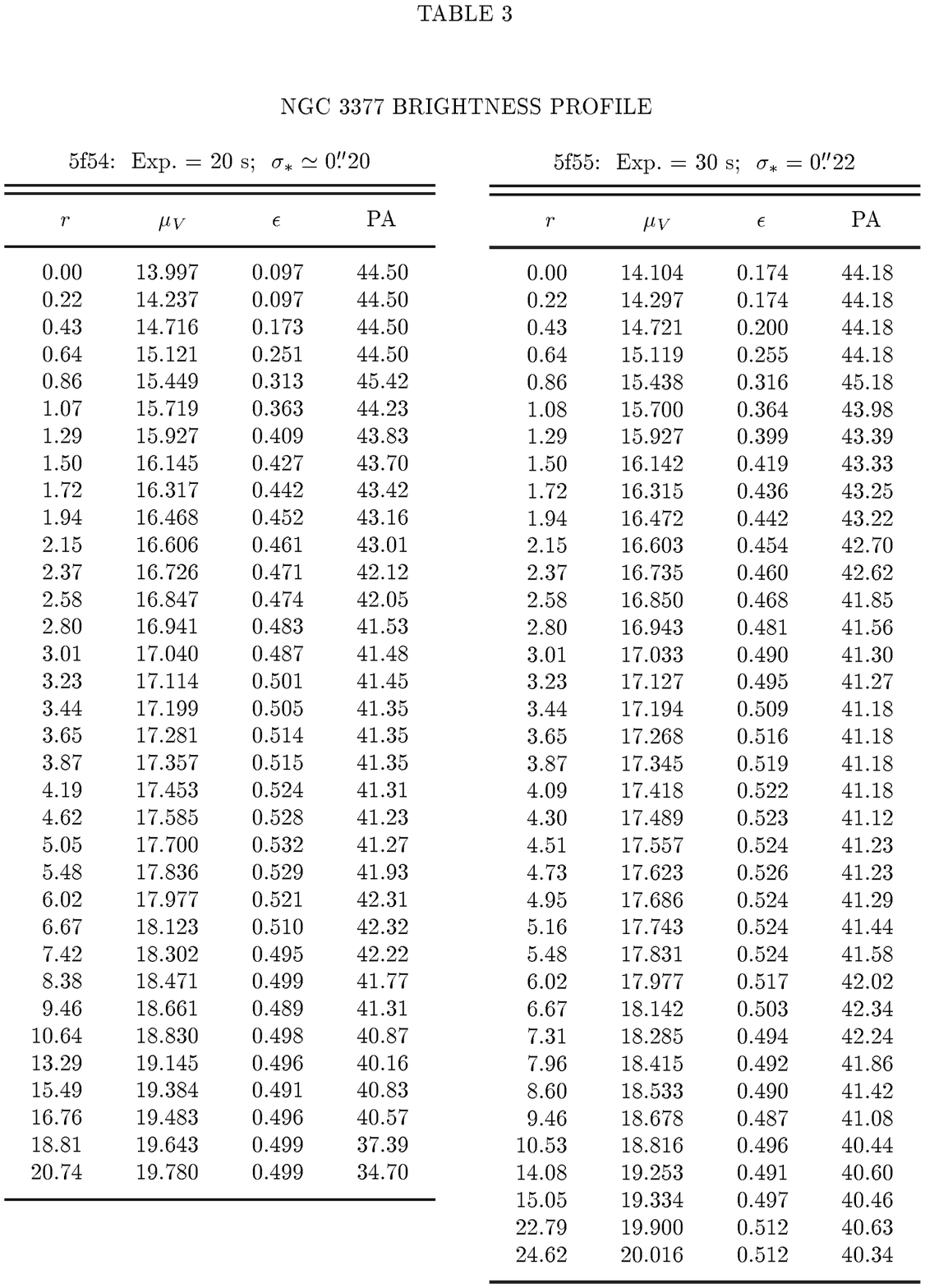}

\eject

\cl{\null} \vfill

\includegraphics{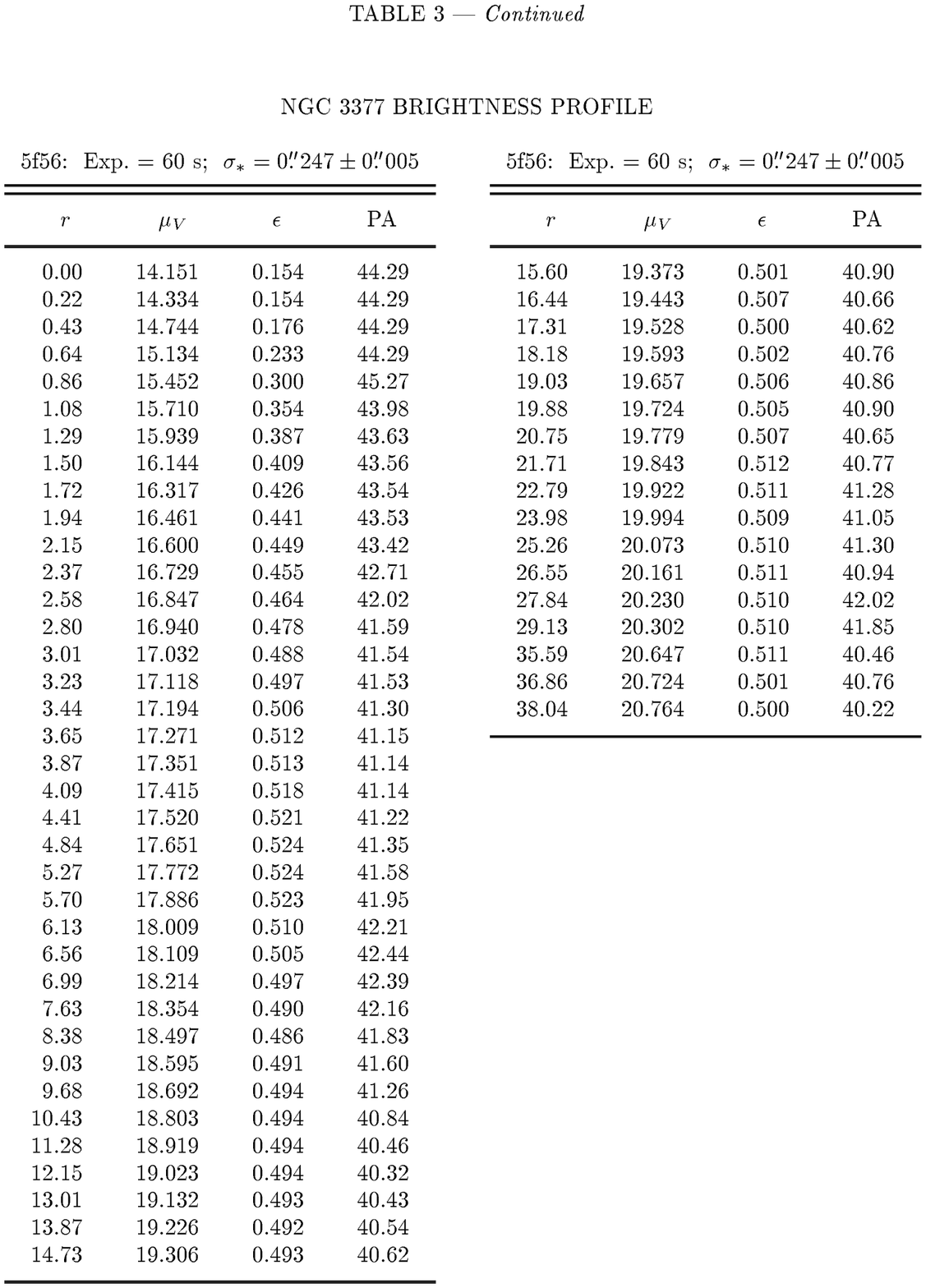}

\eject

\cl{\null} \vfill

\includegraphics{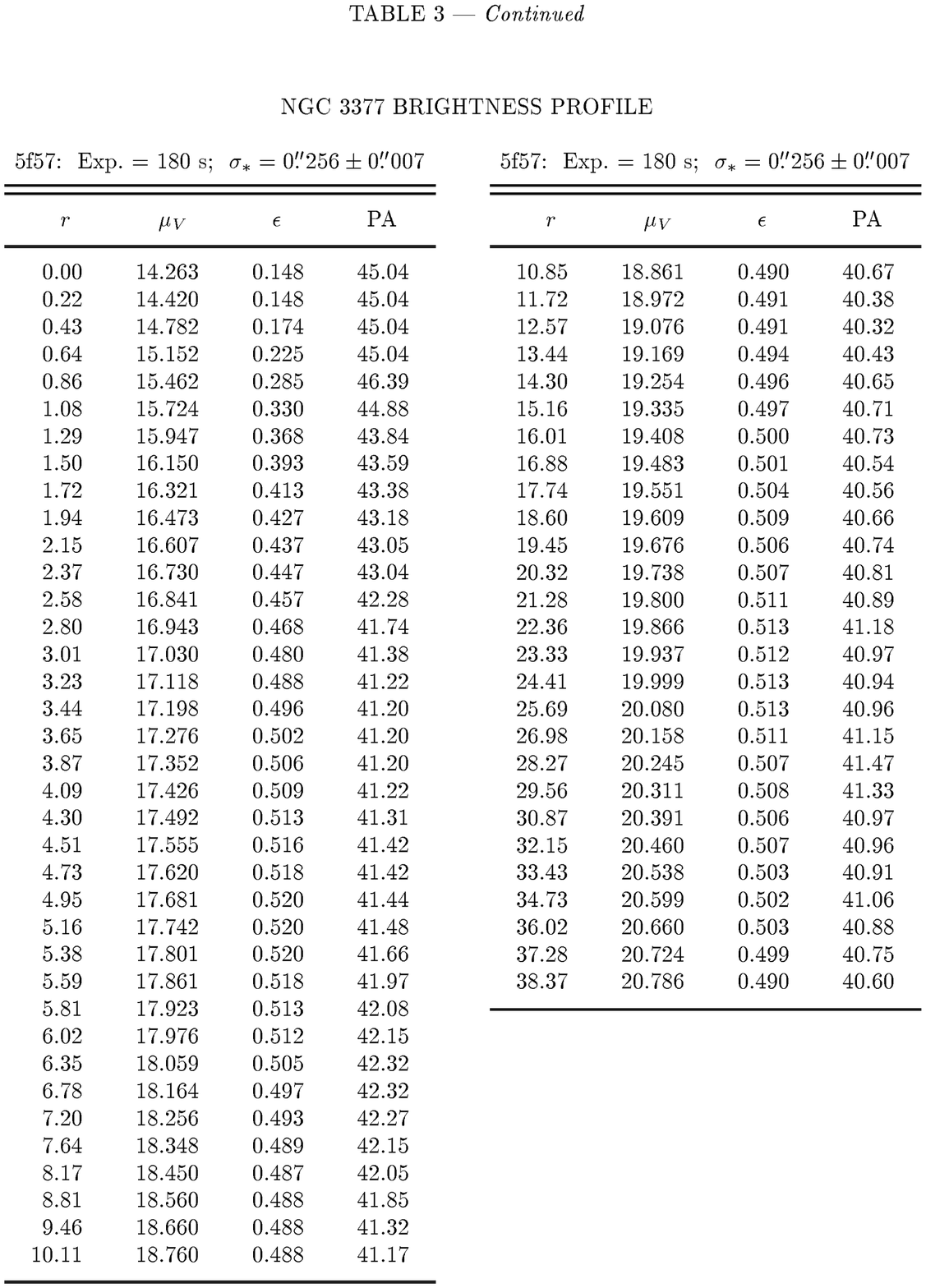}

\eject

\doublecolumns

\cl{\null} \vskip 12.5truecm

\includegraphics{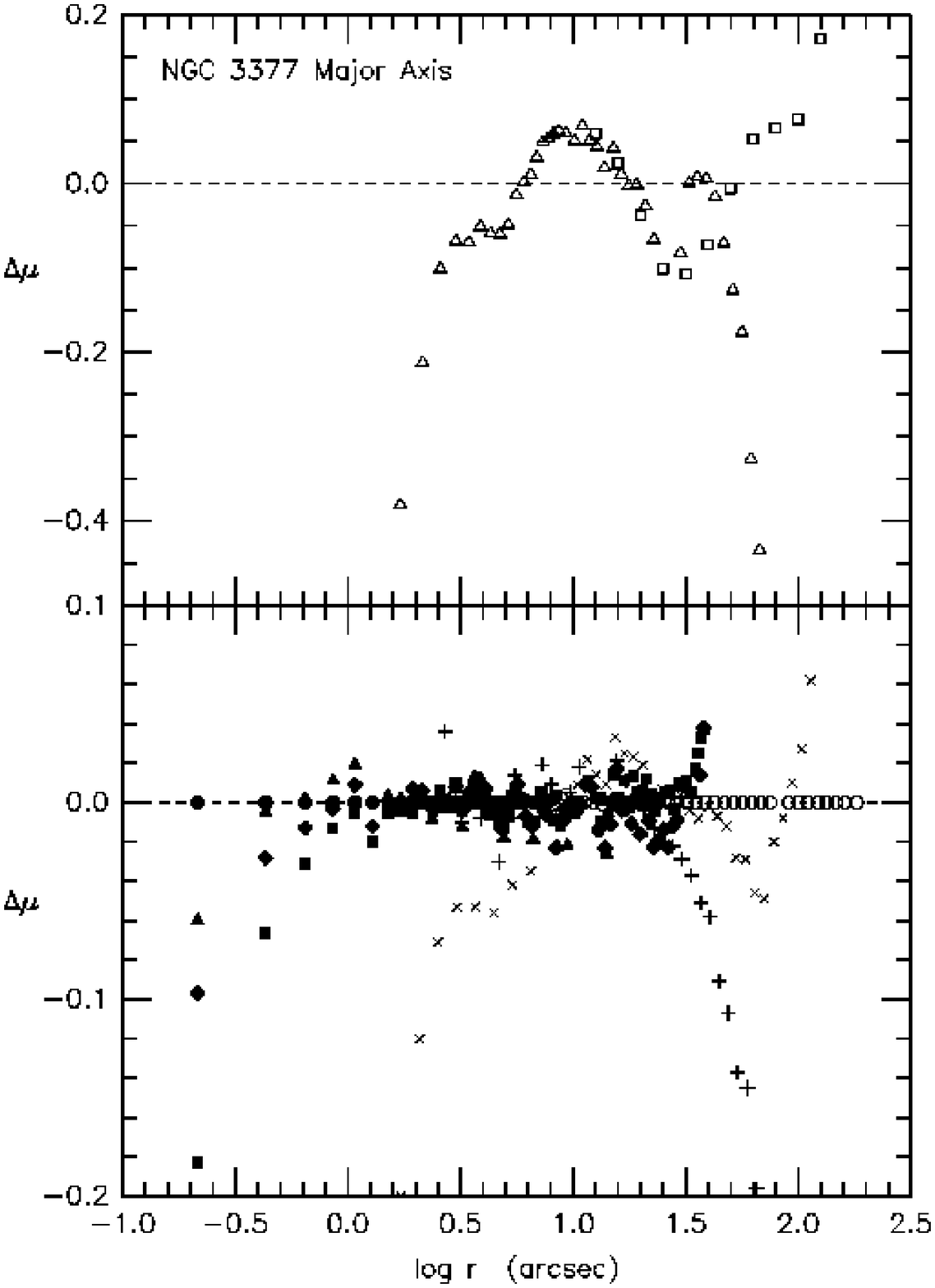}

{\medbaselines\mc{F{\sc IG}.~4.}---Deviations $\Delta\mu$ (mag arcsec$^{-2}$) of
the individual profiles in Figure 3 from the 5f54 profile at small $r$ and the
Pierce (1991) profile at large $r$.  The lower panel shows the most accurate 
photometry; the upper panel shows the rest of the data.  The symbols are the 
same as in Figure 3 except that open triangles are $r$-band data from Djorgovski
(1985).  Near the center, points ``peel away'' from $\Delta\mu = 0$ mag 
arcsec$^{-2}$ where profiles have lower spatial resolution than $\sigma_* = 
0\sd20$.  Deviations at large radii are assumed to be due to inaccurate sky 
subtraction.}

\vss\vskip 8pt

      In \S{\ts}5, we derive volume brightness models that fit these data after
projection and seeing convolution.  To illustrate the precision we should aim
for, we show in Figure 4 the deviations of the photometry from an average of the
5f54 profile at small $r$ and the Pierce profile at large $r$.  This time, all
of the above photometry is included, so deviations at small and large radii are 
apparent.  The upper panel shows the Michard \& Simien (1988) data and 
photometry by Djorgovski (1985).  The latter were derived with a CCD that is 
poor by modern standards (it required large flat-field corrections).  The 
results are less accurate than those in the bottom panel, but they confirm our 
composite profile.  The best data are shown in the bottom panel.  They are 
remarkably consistent.  Except at small and large radii, almost all of the 
scatter is less than $\pm 0.02$ mag arcsec$^{-2}$.  Systematic fitting errors
much larger than this imply that a model disagrees with the photometry.

\huge

\huge

      Finally, Figure 5 shows the ellipticity profile.  Because of the disky
distortion, the apparent ellipticity gets larger as the resolution improves.
The VISTA isophote fitting program is especially sensitive to an edge-on
embedded disk, so the CFHT and {\it HST\/} data show it very strongly.  
As found already by Scorza \& Bender (1995), Figure 5 shows that the embedded
disk contributes most at $r \simeq 5^{\prime\prime}$.  Our modeling results are 
insensitive to the adopted flattening, so we assume that $\epsilon = 0.5$ at
all radii.

\vskip 6.8truecm

\includegraphics{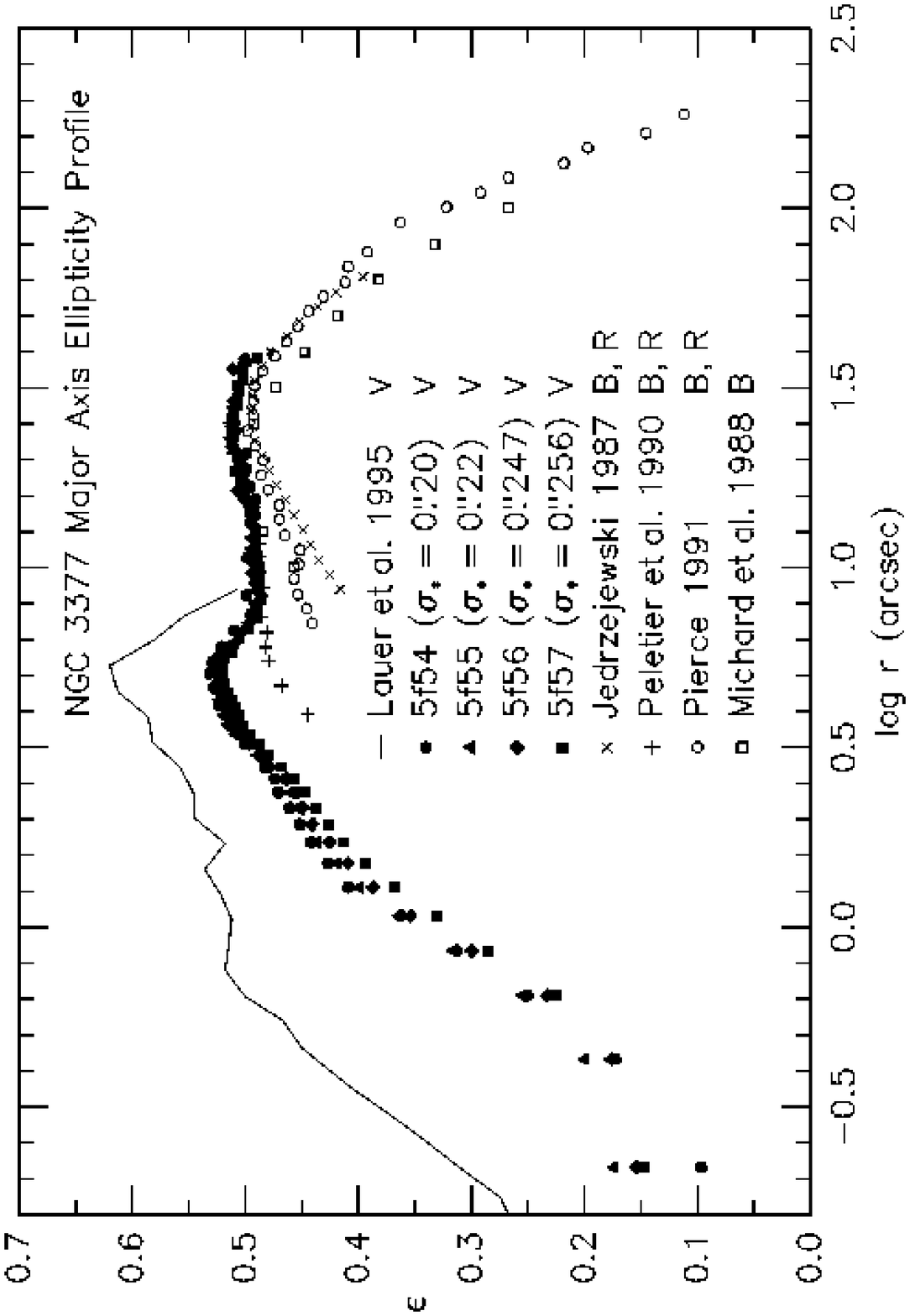}

{\medbaselines\mc{F{\sc IG}.~5.}---Isophote ellipticity $\epsilon = 1 - b/a$ as
a function of major-axis radius in NGC 3377 ($b/a$ is isophote axial ratio).}

\vss\vskip 9pt
\cl {4.~\sc ANALYSIS TECHNIQUE}
\vss

      To derive masses and mass-to-light ratios as a function of radius, we need
seeing-corrected and unprojected brightnesses, rotation velocities, and velocity
dispersions.  As in previous papers, we do not try to invent a deconvolution and
deprojection technique that is powerful enough to produce unique results.
Instead, we convolve models with seeing and compare them to the data.  Also, we
do not try to prove uniqueness.  Rather, we construct fits that bracket the 
observations in $V(r)$, and $\sigma(r)$.  In particular, we find low-mass 
``error bar'' models in which the calculated $V(r)$ and $\sigma(r)$ are too 
small near the center.

      One more piece of machinery is needed.  Fourier quotient measurements
respond nonlinearly to mixtures of stellar populations with different 
dispersions.  Also, the observed dispersion comes partly from rotational line
broadening.  Model calculations therefore mimic the construction by seeing and
projection of the observed spectra.  The first step is to make a library of 
input spectra suitable for NGC 3377.  It consists of the spectrum of the 
standard star broadened to $\sigma = 60$ -- 240 \kms~in steps of 20 \kms; for
each $\sigma$ there are entries at $V = -1000$ to 1000 \kms~in steps of 20 \kms.
The library is used to construct synthetic ``observed'' spectra, as follows. 
We begin with a trial unprojected rotation curve $V(r)$ and dispersion profile 
$\sigma(r)$, both assumed to be independent of distance from the equatorial 
plane.  Then for each radius $r$ along the major axis, consider all other pixels
at radius $\vec r^{\ts\prime}$ and depth $z^{\prime}$ along the line of sight.
We calculate the luminosity-weighted projected spectrum at $\vec r^{\ts\prime}$.
This scatters light into the model pixel by an amount proportional to
$I_*(r_*)$, where $I_*$ is a star profile with $\sigma_* = 0\sd20$ and $r_* = 
|\vec r^{\ts\prime} - \vec r\ts|$.  

\huge

\huge

\ni Here $I_*$ was measured for stars in the bracketing direct images that were
taken~to check the galaxy centering for spectrum 80f027; they were fitted with a
modified Moffat (1969) function to give the PSF used in the analysis,
$$ I_*(r_*) = {1 \over {\biggl[1 + \displaystyle
   \biggl({r_* \over 0{\rlap{.}''\kern -.04em 29701}}\biggr)^{\b\b\lower 3pt\hbox{$\scriptstyle 2.68487$}}\ts\biggr]^{\b\lower 3pt\hbox{$\scriptstyle 1.54730$}}}}~.\eqno(1)$$
The output spectrum is the sum over all scattering pixels of the projected 
spectra weighted by the product of the projected brightness and the star 
profile.  The pixels used in the calculation are smaller by a factor of 3.5 than
the ones in spectrum 80f027 (1 pixel $=$ 0\sd043).  The integration is carried
out to $r_* = 5\sigma_*$ to include the non-Gaussian wings in the star profile.
Finally, the model spectrum is analyzed with the Fourier program.  The
parameters of the model are then varied until the results fit the kinematics.
Uncertainties are estimated by exploring the range of $V(r)$ and $\sigma(r)$
allowed by the data.  This technique was developed independently by Kormendy
(1988a, b) and by Dressler \& Richstone (1988).

      The calculations are routine but time-consuming.  To keep them manageable,
seeing effects are calculated only $|z| = 5\sd1$ deep along the line of sight.
A separate model run with $\sigma_* = 0\s$ does the projection integral for
$5\sd1 \leq |z| \leq 166\s$.  We checked that seeing is not important for $|z| >
5\sd1$.  At each radius in the output image, the two synthetic spectra are sums 
of 116,040 and 3,740 spectra, respectively.  Their sum is the required model 
spectrum.

      Finally, the mass inside $r$ is given by the first velocity moment of the 
collisionless Boltzmann equation,
$$ M(r) = {{V^2r}\over G} + {{\sigma_r^2r}\over 
G}~\biggl[- \ts{{d\ln{\rho}}\over{d\ln{r}}} - 
{{d\ln{\sigma_r^2}}\over{d\ln{r}}} ~~~~~~~~~~~~~~~~~~~~~~~~\null$$

\null \vskip -35pt \null

$$ \null~~~~~~~~~~~~~~~~~~~~~~~~~~~~~~
- \biggl(1 - {\sigma_{\theta}^2 \over \sigma_r^2}\biggr) 
- \biggl(1 - {\sigma_{\phi}^2 \over \sigma_r^2}\biggr)\biggr]. \eqno (2) $$
Here $\sigma_r$, $\sigma_{\theta}$, and $\sigma_{\phi}$ are the radial and two
tangential components of the unprojected velocity dispersion.  Also, $\rho$ is
the unprojected density of the stars that contribute to the spectra.  We assume
that $M/L_V$ for these stars is independent of radius; then $d \ln {\rho} / d 
\ln {r} = d \ln {I} / d \ln {r}$.  Equation (2) is based on the approximations
that the mean rotation is circular and that the mass distribution is spherical.
We will correct the results for the flattening of the galaxy in \S\ts8.

      If $M/L_V$ is nearly constant at large radii and if it rises rapidly 
toward the center at $r$ \lapprox \ts1$^{\prime\prime}$, then there is evidence
for a central dark object.

\vss\vsss\vsss
\cl {5.~\sc VOLUME LUMINOSITY PROFILE}
\vss

      Previous BH papers were based on ground-based measurements of brightness
profiles.  Consequently it was necessary to model the effects of seeing on
$I(r)$ in the same way that we model its effects of $V$ and $\sigma$: we
constructed a series of analytic approximations to $I(r)$ near the center that
bracketed the observed profile after projection and seeing-convolution.  This is
not necessary for NGC 3377, because {\it Hubble Space Telescope} ({\it HST\/})
measurements of its projected profile (Lauer \etal 1995, 1997) have essentially
infinite resolution compared to the spectroscopic resolution.  We have therefore
constructed a composite projected brightness profile from 0\sd022 to 
182$^{\prime\prime}$ by using the {\it HST\/} profile at $r < 4^{\prime\prime}$
and the best ground-based measurements in Figure 3 at $r > 4^{\prime\prime}$.
This profile was deprojected and the result was used as the luminosity model in
all calculations discussed below.  Preliminary results reported earlier 
(e.{\ts}g., Kormendy 1992b) were based on analytic luminosity models such as 
those discussed above; the results did not change significantly when the 
{\it HST\/} profile was adopted because it is almost identical to one of the 
analytic models.

      The mean projected and unprojected major-axis brightness profiles of
NGC 3377 are illustrated in Fig.~6.

\vskip 6.72truecm

\vfill

\includegraphics{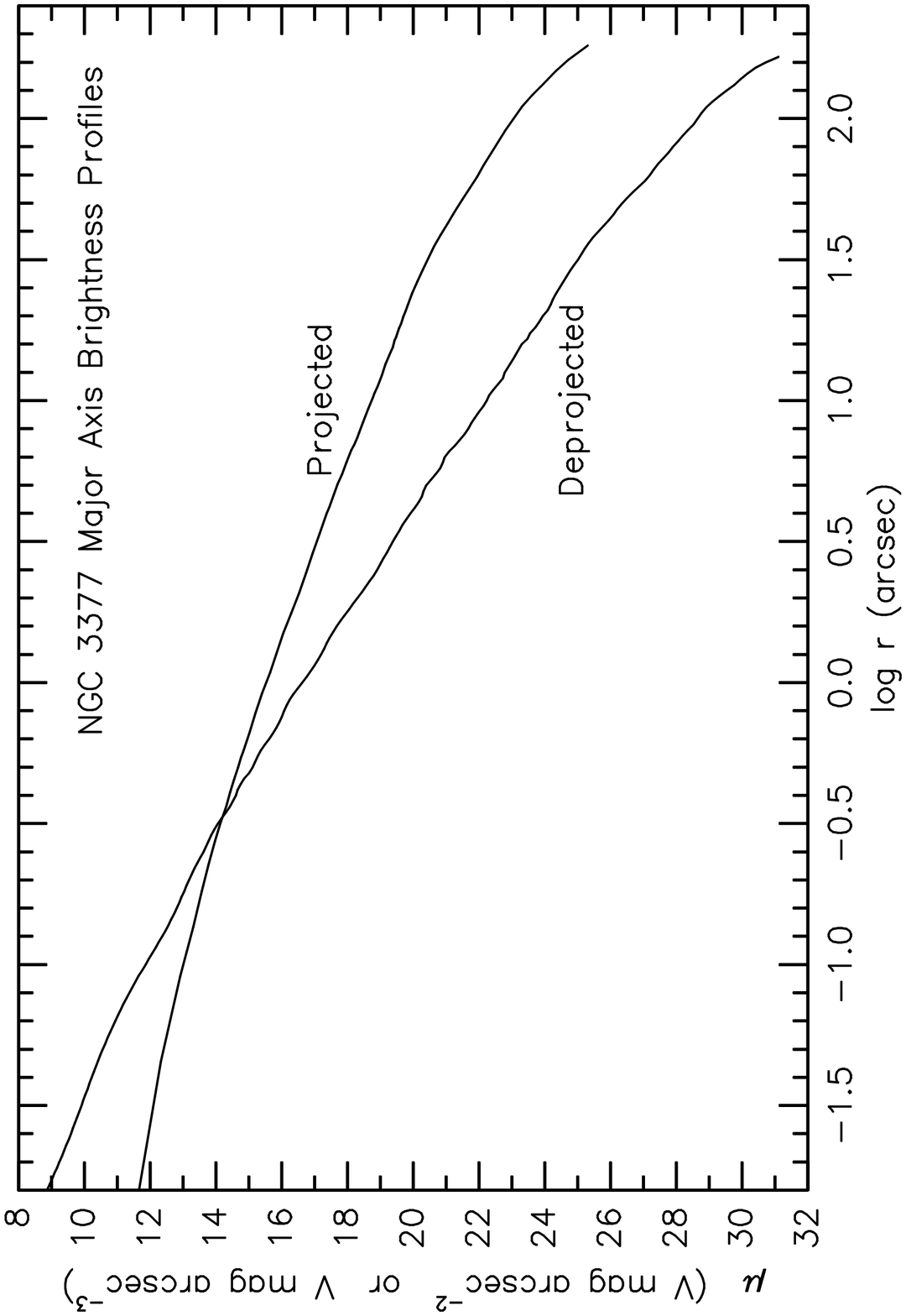}

{\medbaselines\mc{F{\sc IG}.~6.}---Composite {\mit HST\/} and ground-based
major-axis brightness profile of NGC 3377 before and after deprojection.}

\vss\vsss\vsss\vsss\vsss\vsss
\cl {6.~\sc ISOTROPIC KINEMATIC MODELS}
\vss

       We used the machinery of \S\ts4 to find the best-fitting isotropic model 
and several that bracket the observations.  The unprojected model rotation and
dispersion profiles are shown in Figure 7, fits to the kinematic data are shown
in Figure 8, and the mass-to-light ratio profiles are shown in Figure 9.  
Figure 10 shows residual mass-to-light ratios after various MDO masses are
subtracted.  Table 4 lists the model parameters. 

      Unprojected rotation and dispersion profiles were chosen as in previous
papers.  Each rotation curve is the sum in quadrature of a Keplerian and three
rotation curves for exponential disks (Freeman 1970).  This sounds complicated,
but we emphasize that these are no more than convenient fitting functions.  The
total unprojected rotation curves (Fig.~7) are simple: they rise very slowly 
from large radii toward the center and then either drop to zero, stay constant,
or rise steeply inside 1$^{\prime\prime}$.  The Keplerian and the central 
exponential rotation curve are varied to bracket the observed amount of rotation
at $r$ \lapprox \ts1$^{\prime\prime}$ and hence to see how much central mass is
implied by the observations.  Two more exponential disk rotation curves are 
needed to fit the outer rotation curve; the reason is that these are not 
particularly suitable ``basis functions'' to fit an almost-flat rotation curve.
The present models are slightly more complicated than the ones used in earlier
papers because we need to model an almost-flat rotation curve over a larger 
range in radius.  Similarly, the unprojected velocity dispersion is assumed to
be the sum in quadrature of a ``Keplerian'' $\sigma = \sigma_K/r^{1/2}$, $r$ in
arcsec, and a constant $\sigma_c$.  The total $V$ and $\sigma$ are restricted 
to be $\geq 111$ \kms\ and $\geq 70$ \kms, respectively.

\eject

\singlecolumn
\vsize=24.7truecm  \voffset=-0.6truecm


\vskip 12.37truecm


\includegraphics{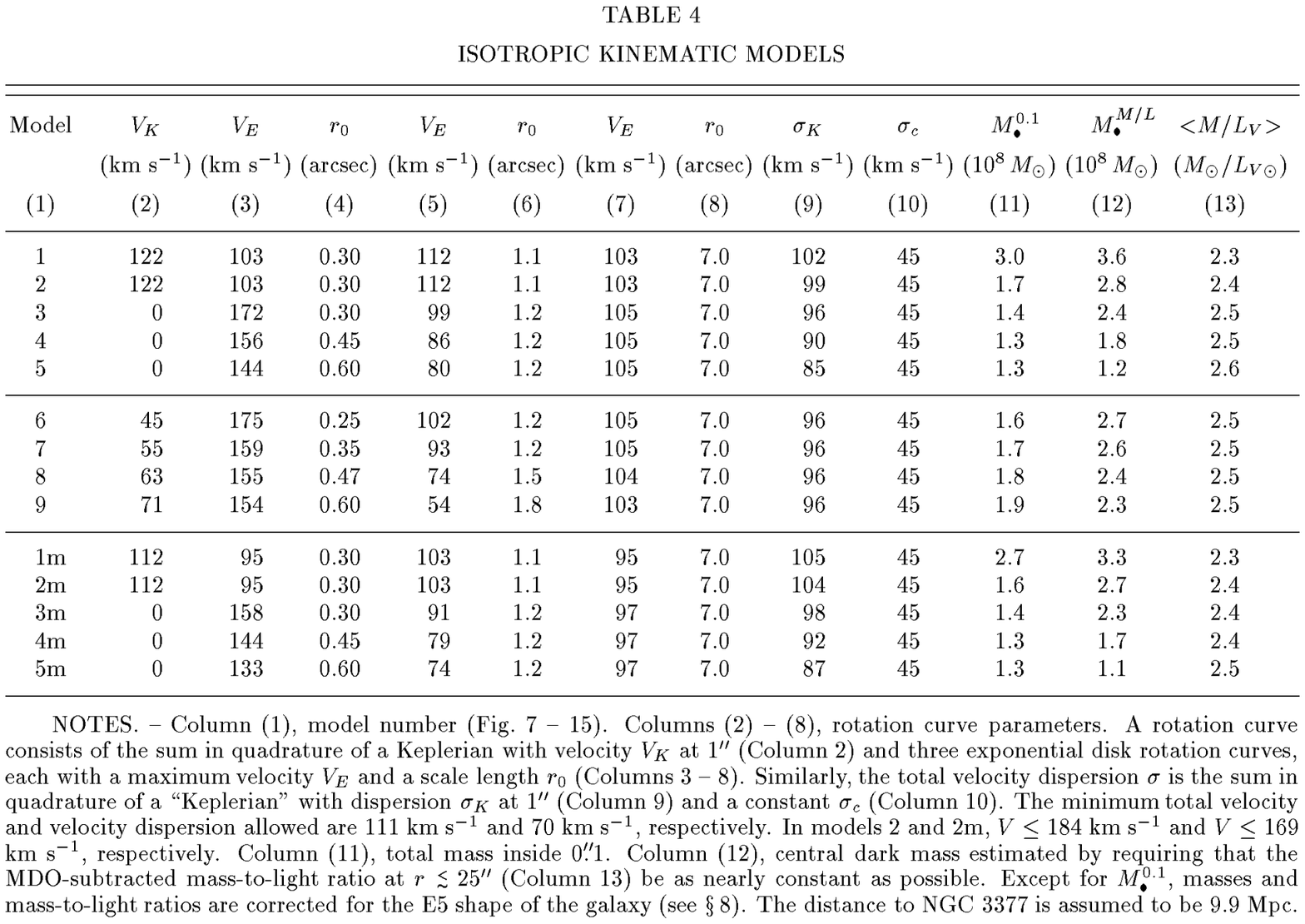}

\doublecolumns

      The best-fitting model, number 3, is a good fit to both the rotation and
the dispersion profiles (Fig.~8).  Its dispersion gradient is not quite steep
enough to be a perfect fit to the low observed $\sigma \simeq 100$ \kms\ at
$1^{\prime\prime}$, and similarly, the rotation curve rises to a slightly 
sharper peak an 0\sd9.  These features in the data are probably due to the 
embedded disky structure (Scorza \& Bender 1995).  But the slope of the inner
rotation curve is fitted essentially perfectly.  If we slightly underestimate 
the slope in $\sigma(r)$, we only underestimate \mbh.  Compared to model 3, 
models 2 and 1 have progressively more rotation and steeper dispersion gradients
near the center.  Model 2 is a reasonable ``error bar'', while model 1 is 
clearly excluded by the data.  Similarly, models 4 and 5 have too little 
rotation and dispersion gradient near the center; model 4 is a reasonable 
``error bar'', while model 5 is clearly excluded.

      Figure 9 shows the corresponding mass-to-light ratios.  It is interesting
and reassuring that all models imply constant mass-to-light ratios $M/L_V \simeq
2.6$ to 3 between $r \simeq 2\s$ and 35$^{\prime\prime}$.  This means 
that the inner part of the galaxy is dominated by an old stellar population, 
with no significant contribution from halo dark matter.  The same is true in NGC
3115 (Kormendy \& Richstone 1992) and in NGC 4594 (Kormendy \& Westpfahl 1989). 
In all three galaxies, the $M/L$ profile at large radii is simple and requires 
no unseen mass other than that normally associated with stars.  This is a useful
result in its own right.  Since the volume brightness changes by a factor of
$\sim 630$ over this radius range, and since the modeling machinery allows 
$M/L_V$ to vary as much as it likes, the constancy of $M/L_V$ is a good sign 
that the machinery is working correctly.

      In contrast, $M/L_V$ increases at $r < 2\s$ by a factor of at least 4.
The absolute values to which $M/L_V$ rises are not larger than we see in more
luminous ellipticals.  Also, the rise is important only at small radii. Finally,
it is possible to make anisotropic, three-integral maximum entropy models 
(Richstone\ts{\it et al.}\ts1998a) that fit the data without an MDO.  Therefore
this is a weaker MDO detection than those in the Galaxy, M{\ts}31, M{\ts}32, and
NGC 3115 (KR95).

      As in previous papers, we derive two estimates of the mass of the central
dark object (Table 4).  First, $M_{\bullet}^{0.1}$ is the total mass interior to
$r = 0\sd1$.  Unlike estimates that depend on the mass distribution at $r$ 
\gapprox \ts0\sd5, this value is not corrected for the flattening of the 
galaxy, because the potential near the center is almost spherical.  If the 
stellar mass-to-light ratio is nearly constant with radius as suggested by the 
lack of strong line-strength gradients, then Fig.~9 implies that essentially all
of the mass inside 0\sd1 radius is dark.  However, $M_{\bullet}^{0.1}$ is very
sensitive to the effects of projection and seeing.  Also, when it is subtracted
from the mass distribution, a mass-to-light ratio gradient remains.  A second
estimate is therefore derived by requiring that the residual $M/L_V(r)$ be as
nearly constant as possible after a central point mass 
\lower 1pt\hbox{$M_{\bullet}^{M/L}$} is 
subtracted.~~~Figure 10 shows the residual $M/L_V(r)$.  It varies slightly with 
radius; the MDO mass is therefore uncertain by about $\pm 0.4 \times 10^8$
\msun (slightly more for model 1 and less for model 5). The adopted mass for
\hbox{model 3} is \lower 1pt\hbox{$M_{\bullet}^{M/L}$} = $2.4 \times 10^8$ 
\msund.  The ``error bar'' models 2 and 4 imply that the uncertainty is about
$\pm 0.7 \times 10^8$ \msund. 

\vfill\eject

\singlecolumn

\vsize=24.3truecm  \voffset=-0.6truecm

\cl{\null} \vfill

\includegraphics{./3377fig7.cps}

{\medbaselines\mc{F{\sc IG}.~7.}---Intrinsic (i.{\ts}e., not projected or
seeing-convolved) rotation and dispersion profiles for kinematic models 1 -- 5
in Table 4.}

\cl{\null} 
\cl{\null} 
\cl{\null} \vfill

\includegraphics{./3377fig8.cps}

{\medbaselines\mc{F{\sc IG}.~8.}---Models 1 -- 5 (top to bottom at 0\sd4) 
fitted to the major-axis kinematic data.  Here the models have been projected
and convolved with seeing as discussed in \S\ts4.  Only the high-resolution 
data are plotted inside $r = 2^{\prime\prime}$.}

\eject

\cl{\null} \vskip 7.8truecm

\includegraphics{./3377fig9.cps}

{\medbaselines\mc{F{\sc IG}.~9.}---Mass-to-light ratio {\mit M/L}$_V$ (solar
units) as a function of radius for models 1 -- 5, corrected for the E5
shape of the galaxy.} 
\vss

\doublecolumns

\ni Averaging this result with $M_{\bullet}^{0.1}$, we arrive at our final 
estimate for the MDO mass given by models 1 -- 5, \mbh\ = $(1.9 \pm 0.8) \times
10^8$ \msund.  The errors are not Gaussian: a range of \mbh\ values is allowed
because of parameter coupling in the models, but outside the above range, masses
become rapidly excluded by the present machinery.

      Model 3 fits the data and implies that NGC 3377 contains an MDO, but it
has a shortcoming.  The rotation velocity decreases to zero at the center.  This
is not impossible in principle -- the BH detection comes mostly from the 
dispersion gradient -- but it is not the natural expectation.  It is important
to note that we are just measuring masses, like when we use an embedded H{\ts}I
disk to measure the mass of a galaxy, but with the added complication of a 
non-zero velocity dispersion.  We are not making self-consistent dynamical 
models.  So we may be allowing some freedom in the tradeoff between $V$ and 
$\sigma$ that the galaxy does not have.  The results of such models have proved
to be very reliable in the past (see, e.{\ts}g., Kormendy \etal 1996a,{\ts}b and
van der Marel \etal 1997a,{\ts}b, which compare {\it HST\/} results on NGC 3115,
NGC 4594, and M{\ts}32 with BH results from papers using techniques like ours).
Nevertheless, the tradeoff between $V$ and $\sigma$ deserves further 
exploration.

      We have therefore constructed a more closely spaced series of models
that have various small amounts of Keplerian rotation curve near the center.
These are models 6 -- 9 in Table 4; they are illustrated in Figures 11 -- 13.
Models 7 and 8 are good fits to the data.  In fact, they are better fits than
model 3, because the extra rotational line broadening allows a better fit of the
central velocity dispersion.  Models 6 and 9 are error bars to $V(r)$, although
they are not excluded by the data.  Their rotation curves are slightly more
complicated that those of models 1 -- 5 (see Fig.~11).  Other, similar models
are possible.  But these, too, have steeply rising mass-to-light ratio profiles
near the center (Fig.~13), and in these cases, the central dark mass is implied
by both $V$ and $\sigma$.

      Together, models 1 -- 9 show that the detection of a central dark object
in NGC 3377 is very robust, within the assumption that the velocity distribution
is isotropic.  The implied mass is \mbh\ $= (1.9 \pm 0.8) \times 10^8$ \msund.
It requires two small corrections.   It already contains a correction for the 
flattening of the galaxy; this is discussed in \S\ts8.  Also, in \S\ts7, we 
correct for the fact that we fit Gaussians to the line-of-sight velocity 
distributions instead of measuring moments as required by Equation (2). 

\vskip 120.4truemm

\includegraphics{./3377fig11.cps}

{\medbaselines\mc{F{\sc IG}.~11.}---Intrinsic (i.{\ts}e., not projected or
seeing-convolved) rotation and dispersion profiles for kinematic models 6 -- 9.}

\eject

\singlecolumn

\cl{\null} \vfill

\includegraphics{./3377fig10.cps}

{\medbaselines\mc{F{\sc IG}.~10.}---Mass-to-light ratio {\mit M/L}$_V$ as a 
function of radius for models 1 -- 5 before (heavy solid lines) and after (light
lines) subtraction of the MDO masses listed in the keys.  At $r < 0\sd2$, the
curves get very noisy and so are not plotted.  For each model, residuals are 
shown for four values of \lower 1.5pt\hbox{{\mit M$_\bullet^{{\ts}M/L}$}}; the
second-smallest, which corresponds to the light solid line, is the adopted
value.  The others (dashed lines) illustrate the uncertainty that results from
the fact that {\mit M/L$_V$} is not exactly constant with radius.  The
horizontal straight line in each panel is the adopted mean stellar mass-to-light
ratio.  All mass-to-light ratios are corrected for the E5 shape of the galaxy.}

\eject

\doublecolumns

\cl{\null} \vskip 117.0truemm

\includegraphics{./3377fig12.cps}

{\medbaselines\mc{F{\sc IG}.~12.}---Models 6 -- 9 (top to bottom at 0\sd6) 
fitted to the major-axis kinematic data.  Here the models have been projected
and convolved with seeing.  Only the high-resolution data are plotted.}

\cl{\null} \vskip 93.0truemm

\includegraphics{./3377fig13.cps}

{\medbaselines\mc{F{\sc IG}.~13.}---Mass-to-light ratio {\mit M/L}$_V$ (solar
units) as a function of radius for models 6 -- 9, corrected for the E5 
flattening of NGC 3377 as discussed in \S\ts8.} 
\vss

\vss\vsss\vsss
\cl {7.~LINE-OF-SIGHT VELOCITY DISTRIBUTIONS}
\vs

      The quantities $V$ and $\sigma$ in Equation (2) are moments of the
line-of-sight velocity distribution (LOSVD).  In real life, moments cannot be
measured, because they are sensitive to small numbers of stars that are far from
the mean velocity and hence out in the (unknown) continuum.  In \S\ts2, we 
derived velocities and velocity dispersions by fitting Gaussians to the line
profiles.  Several authors have pointed out that the derivation of $M(r)$ can 
suffer systematic errors if such measurements are substituted for moments (van
der Marel \& Franx 1993; van der Marel \etal 1994a,{\ts}b; Bender {\it et al.}
1994).  However, KR95 (see p.~598 -- 599) point out that the moment $V(r)$ is 
proportional to the Gaussian fit $V(r)$, so the main effect is to lower the 
global (i.{\ts}e., bulge) mass-to-light ratio slightly; the strength of the BH
case is almost unchanged.  Since many BH papers are based on Gaussian fit $V$
and $\sigma$ values, it is useful to illustrate this point for at least one 
galaxy.  Therefore we investigate here how LOSVD asymmetries affect the 
conclusions of \S\ts6.

      We measured non-parametric LOSVDs using the Fourier Correlation Quotient
method (FCQ, Bender 1990).  To extract higher-order information on the LOSVD
shapes, we fitted them with a Gaussian plus third- and fourth-order 
Gauss-Hermite polynomials $H_3$ and $H_4$ (van der Marel \& Franx 1993, Gerhard 
1993): 
$${\rm LOSVD}(v) = {\gamma \over \sqrt{2\pi\sigma^2}}\; e^{-{(v-V)^2 \over
   2\sigma^2}}~.\quad\quad\quad\quad\quad\quad\quad\quad\quad\phantom{0}$$
$$\phantom{0}\quad\quad\quad\quad .~\biggl[1 + h_3 H_3\biggl({v-V\over 
  \sigma}\biggr) + h_4 H_4\biggl({v-V\over \sigma}\biggr)\biggr]~. \eqno{(3)} $$
\vs
\ni The coefficients $h_3$ and $h_4$ parametrize the lowest-order asymmetric and
symmetric deviations from Gaussian line profiles.  Positive $h_4$ implies that
the line profile is more triangular than Gaussian; i.{\ts}e., it is more 
strongly peaked and has more extended wings than the best-fitting Gaussian.  
Negative $h_4$ parametrizes deviations toward rectangular line profiles.
The $h_4$ amplitudes are generally small in ellipticals, with values less than
a few percent (Bender \etal 1994).  This is also the case in NGC 3377: $h_4
\simeq 0$ at $r \leq 6^{\prime\prime}$.  The FCQ results are shown in Fig.~14.

    The $h_3$ amplitude can reach values of $\pm 0.15$; it couples with rotation
as $h_3 \simeq -0.1{\ts}V/\sigma$ (Bender {\it et al.}~1994).  In NGC 3377,
$h_3 = -0.09{\ts}V/\sigma$ with no significant deviations (Fig.~14). 
Therefore there are more stars on the retrograde (systemic-velocity) side of the
LOSVD than on the prograde side.  Such behavior is natural for rotating stellar
systems (see the above references); it is especially expected when there is a 
disky, rapidly-rotating structure embedded in a more nearly spherical and 
slowly-rotating body (Scorza \& Bender 1995).

      Having measured $h_3$ and $h_4$ profiles, we can derive more realistic 
estimates of the velocity moments than we got from Gaussian fits alone.  For
the case $h_4 \simeq 0$ and $-0.15 < h_3 < 0$, Bender \etal (1994, see Fig.~3)
find that \vskip -8pt

$$\eqalignno{{V_{\rm mom} - V_{\rm fit} \over \sigma_{\rm fit}}\; &=\; 
                                          -1.0{\ts}|h_3|^{0.83}~; &(4)\cr
                                                                      \cr
  {\sigma_{\rm mom} - \sigma_{\rm fit} \over \sigma_{\rm fit}}\; &=\;
                          \phantom{-}4.7{\ts}|h_3|^{2.4}~.        &(5)\cr} $$

\ni Here $V_{\rm mom}$ and $\sigma_{\rm mom}$ are the approximate true moments,
and $V_{\rm fit}$ and $\sigma_{\rm fit}$ are derived by fitting Equation (3) to
the LOSVDs.  Moments derived in this way are illustrated in Figure 14, together 
with the velocities and velocity dispersions given by the Fourier Quotient and 
Fourier Correlation Quotient methods.  The velocity dispersions are virtually 
identical for all three procedures, but the velocities differ by up to 15\ts\%. 
As expected for negative $h_3$, the velocity moments are smaller than both the
FQ and FCQ fitted values.  However, the FQ values are between the FCQ and moment
velocities.  Therefore Gaussian fits give velocities that are closer to the true
moments than higher-order fits incorporating $h_3$ and $h_4$.

\vskip 138truemm

\includegraphics{./3377fig14.cps}

{\medbaselines\mc{F{\sc IG}.~14.}---Fourier Correlation Quotient measurements
of $V$, $\sigma$, $h_3$, and $h_4$ along the major axis of NGC 3377.  The
FQ results are shown for comparison.  In the $h_3$ panel, the lines show $h_3 =
-0.09{\ts}V/\sigma$ calculated from the filled circles in the bottom two
panels.} \vss

      How do these results affect the mass measurements in this and previous 
BH papers?  Figure 14 confirms the remark in KR95 that the velocity moments are
smaller than the FQ velocities by a scale factor that is almost independent of
radius.  Models 1m -- 5m (Table 4) are models 1 -- 5 recalculated with 
velocities scaled down by a factor of 0.92.  They fit the moment data in Figure
14 in the same way that models 1 -- 5 fit the FQ data in Figure 8.  Model 3m is
a good fit; model 4m provides a low-\mbh~error bar, and model 5m clearly does
not fit the data.  Models 1m and 2m similarly provide high-\mbh~error bars. 
\huge

\huge

\vskip 130.0truemm

\includegraphics{./3377fig15.cps}

{\medbaselines\mc{F{\sc IG}.~15.}---Comparison of Fourier Quotient measurements,
Fourier Correlation Quotient measurements, and approximate velocity moments
calculated from the FCQ measurements using Equations 4 and 5.  The lines show
models 1m -- 5m, i.{\ts}e., models 1 -- 5 with velocities scaled down by a
factor of 0.92.}
\vss

     Figure 16 shows the mass-to-light ratio profiles for models 1m -- 5m.  As
expected, they are similar in shape to those in Fig.~9; i.{\ts}e., $M/L_V$ still
climbs quickly at $r$ \lapprox \ts2$^{\prime\prime}$.  This means that the
strength of the BH case is essentially unchanged.  But the overall mass scale is
slightly smaller.  The factor is not as extreme as 0.92$^2$, because $\sigma(r)$
must be made steeper to preserve a good fit to the data now that rotational line
broadening is reduced.  The BH masses and mass-to-light ratios are given for 
models 1m -- 5m in Table 4. We find that \mbh\ = $(1.8 \pm 0.8) \times 10^8$
\msund.

      Note that models 1m -- 5m are not attempts to fit the full LOSVDs of NGC
3377, i.{\ts}e., they are not intended to model the complete dynamical behavior
of the galaxy.  They have essentially Gaussian LOSVDs; this is why they do not
themselves require a correction from Gaussian fit to moment $V$ and $\sigma$ 
values.  We emphasize that the purpose of this paper is only to measure the mass
distribution well enough to search for a BH.  With models 1m -- 5m, we have
corrected our measurements of $M(r)$ and \mbh~for the most important effects
of LOSVD asymmetries.  As expected, the corrections are small.  Similar
conclusions apply to past BH papers based on FQ measurements: the errors 
made by neglecting $h_3$ and $h_4$ are small compared to other uncertainites
in the analysis.  \hbox{This is especially true}
\huge\vskip -12pt

\cl{\null} \vskip 84.truemm

\includegraphics{./3377fig16.cps}

{\medbaselines\mc{F{\sc IG}.~16.}---Mass-to-light ratio {\mit M/L}$_V$ (solar
units) as a function of radius for models 1m -- 5m, corrected for the E5 
flattening of NGC 3377 as discussed in \S\ts8.} 
\vss\vskip 9pt

\ni when bulge-subtracted spectra were analyzed (Kormendy 1988b,{\ts}c), since
bulge subtraction removes most of the LOSVD asymmetries (Kormendy 1994).

\vss\vsss\vsss\vsss\vsss
\cl {8.~DISCUSSION}
\vs

      Isotropic models imply that NGC 3377 contains a central dark object, 
probably a BH, of mass \mbh\ $\sim 2 \times 10^8$ \msund.  The BH mass and 
stellar mass-to-light ratio require one more correction.  While the fits of the
kinematic models to the observations were correctly based on an E5 light
distribution, the $M(r)$ and $L(r)$ calculations were based on the approximation
that the galaxy is spherical.  For any test particle at radius $r$ along the 
major axis, the observed velocities imply less mass in a flattened galaxy 
because stars are on average closer to the test particle than we assumed. 
Binney \& Tremaine (1987) give the required correction in their Fig.~2-12.
This is for a modified Hubble density distribution, which is not a bad
approximation here.  The E0 and E5 rotation curves are proportional to each 
other; the flattening correction to $M(r)$ is a factor of 0.80.  This is 
approximate, because we assume that the mass distribution is E5 everywhere,
whereas in reality, it gets rounder as $r \rightarrow 0$ because of the BH.  
Nevertheless, the correction should be accurate enough so that other 
uncertainities dominate.  Similarly, the correction to $L(r)$ is a factor of
0.5.  Therefore we correct \mbh~by a factor of 0.80 and $M/L(r)$ by a factor of
$0.80/0.5 = 1.61$.  Table 4 and Figures 9, 13, and 16 include these corrections.

      We conclude that isoptropic kinematic models imply that NGC 3377 contains
a central dark object of mass \mbh~$= (1.8 \pm 0.8) \times 10^8$ \msund.  The
stellar mass-to-light ratio $M/L_V = 2.4 \pm 0.2$ is smaller than normal for an
elliptical of absolute magnitude $M_B = -18.8$ (e.{\ts}g., Kormendy 1987b, 
Fig.~3; note that this is based on a Hubble constant of 50 \kms~Mpc$^{-1}$).
The BH mass supports the emerging correlation between bulge luminosity and 
\mbh~(Fig.~17; Kormendy \etal 1997; KR95; Kormendy 1993).

\cl{\null}
\vskip 7.0 truecm

\includegraphics{./3377fig17.cps}

      {\medbaselines\mc{F{\sc IG}.~17.}---BH mass as a function of bulge 
absolute magnitude, from Kormendy {\mit et al.}~(1997) but with M{\ts}84 (Bower
{\mit et al.}~1998; Green 1997) and NGC 4342 (van den Bosch \& Jaffe 1997;
Cretton \& van den Bosch 1998) added and with NGC 3377 updated.  Circles, 
diamonds, and squares show objects with stellar-dynamical, ionized gas 
dynamical, and maser evidence for BHs, respectively.  Upper limits on \mbh\ are
plotted as crosses.  The correlation may be only the upper envelope of a 
distribution that extends to smaller $M_{\bullet}$.} 
\vss\vskip 8pt

      Our detection of an MDO in NGC 3377 is not quite definitive, because
anisotropic, three-integral maximum entropy models (Richstone \etal 1998a) can 
fit our data without a BH (Richstone \etal 1998b). However, several
arguments suggest that NGC 3377 is not likely to be very radially anisotropic.
To explain the data without a BH requires that $\sigma_r \gg \sigma_\phi$ and 
$\sigma_\theta$.  But NGC 3377 contains an embedded nuclear disk 
(Scorza \& Bender 1995).  Our understanding of how these form -- gas dissipation
and star formation -- would not tend to make $\sigma_r$ very large compared to
$\sigma_\phi$ and $\sigma_\theta$.  In fact, van der Marel \etal (1997b) 
conclude that valid models of M{\ts}32, whose kinematics
are similar to those of NGC 3377, ``are all dominated by azimuthal motion at
most radii'' (their Fig.~10).  Also, NGC 3377 rotates rapidly enough to be near 
the ``oblate line'' in the well known $V/\sigma$ -- $\epsilon$ diagram
(Illingworth 1977; Binney 1978, 1980, 1981, 1982; Binney \& Tremaine 1987; 
$\epsilon =$ ellipticity and $V/\sigma =$ ratio of the maximum rotation velocity
to a mean velocity dispersion near the center).  The oblate line, $V/\sigma 
\simeq [\epsilon/(1 - \epsilon)]^{1/2}$ (Kormendy 1982), describes oblate 
spheroids that are isotropic and flattened only by rotation.  This is a direct 
constraint only on $\sigma_z/(\sigma_r^2 + \sigma_\phi^2)^{1/2}$, but rapidly 
rotating ellipticals have generally turned out to be less anisotropic than 
slowly rotating ellipticals.  Finally, isotropic model measurements of \mbh\ 
have turned out to be close to correct when anisotropic models were constructed
(see KR95 for a review).  So it is likely that our measurement of \mbh\ is
reasonably accurate.  Nevertheless, it will be important to see whether
higher-resolution observations confirm our MDO detection.

      This work is in progress: Richstone \etal (1998b) have obtained 
{\it HST\/} FOS spectroscopy of NGC 3377.  Their three-integral maximum entropy
models of NGC 3377 show conclusively that it contains a central dark object of
mass \mbh\ $\simeq 10^8$ \msund.

\vss\vsss

      JK thanks the staff of the Canada-France-Hawaii Telescope for their 
support of the observing runs.  He is especially grateful to D.~Salmon for his 
care in setting up the f/4 module of the Herzberg Spectrograph.  JK is also 
grateful to N.~Trentham for his assistance during Run 3.  We thank G.~Bower, 
R.~Green, and F.~van den Bosch for allowing us to include their \mbh\ 
measurements in Fig.~17 before publication.  We also thank the referee, R.~van
der Marel, for his helpful comments.  JK and AE were supported in part by NSF 
grants AST-8915021 and AST-9219221.  RB's work was supported by 
Sonderforschungsbereich 375 of the German Science Foundation and by the 
Max-Planck-Gesellschaft.

\huge

\singlecolumn

\vskip -8pt

\vss
\cl{\mc REFERENCES}
\vss\vskip -5pt

\doublecolumns

{\mc\medbaselines


\nhi Bacon, R., Emsellem, E., Monnet, G., \& Nieto, J.~L.~1994, A\&A, 281, 691

\nhi Bender, R.~1988, A\&A, 193, L7

\nhi Bender, R.~1990, A\&A, 229, 441

\nhi Bender, R., D\"obereiner, S., \& M\"ollenhoff, C.~1988, A\&AS, 74, 385

\nhi Bender, R., Kormendy, J., \& Dehnen, W.~1996, ApJ, 464, L123

\nhi Bender, R., Saglia, R.~P., \& Gerhard, O.~E.~1994, MNRAS, 269, 785

\nhi Bender, R., Surma, P., D\"obereiner, S., M\"ollenhoff, C., \& Madejsky,
      R.~1989, \hbox{A\&A,} 217, 35

\nhi Binney, J.~1978, MNRAS, 183, 501

\nhi Binney, J.~1980, MNRAS, 190, 421

\nhi Binney, J.~1981, in The Stucture and Evolution of Normal Galaxies,
      ed.~S.~M.~Fall \& D.~Lynden-Bell (Cambridge: Cambridge Univ.~Press), 55

\nhi Binney, J.~1982, in Morphology and Dynamics of Galaxies, Twelfth Advanced
      Course of the Swiss Society of Astronomy and Astrophysics, ed.~L.~Martinet
      \& M.~Mayor (Sauverny: Geneva Observatory), 1

\nhi Binney, J., \& de Vaucouleurs, G.~1981, MNRAS, 194, 679

\nhi Binney, J., \& Tremaine, S.~1987, Galactic Dynamics (Princeton: Princeton
      Univ. Press)

\nhi Bower, G., {\mit et al.}~1998, ApJL, in press


\nhi Burstein, D., \& Heiles, C.~1984, ApJS, 54, 33

\nhi Caldwell, N.~1983, ApJ, 268, 90



\nhi Carter, D.~1987, ApJ, 312, 514

\nhi Carter, D., \& Jenkins, C.~R.~1993, MNRAS, 263, 1049 



\nhi Cretton, N., \& van den Bosch, F.~C.~1998, in preparation


\nhi Davies, R.~L., Efstathiou, G., Fall, S.~M., Illingworth, G., \&
      Schechter, P.~L.~1983, ApJ, 266, 41

\nhi Dehnen, W.~1995, MNRAS, 274, 919 


\nhi de Vaucouleurs, G.~1948, Ann.~d'Ap., 11, 267






\nhi Djorgovski, S.~B.~1985, PhD Thesis, University of California at Berkeley

\nhi Dressler, A.~1984, ApJ, 286, 97


\nhi Dressler, A., \& Richstone, D.~O.~1988, ApJ, 324, 701

\nhi Emsellem, E., Monnet, G., Bacon, R., \& Nieto, J.-L.~1994, A\&A, 285, 739


\nhi Faber, S.~M., {\mit et al.}~1997, AJ, 114, 1771


\nhi Ferrarese, L., Ford, H.~C., \& Jaffe, W.~1996, ApJ, 470, 444




\nhi Franx, M., Illingworth, G., \& de Zeeuw, T.~1991, ApJ, 383, 112

\nhi Freeman, K.~C.~1970, ApJ, 160, 811



\nhi Gerhard, O.~E.~1993, MNRAS, 265, 213 


\nhi Green, R.~F.~1997, private communication

\nhi Harms, R.~J., {\mit et al.}~1994, ApJ, 435, L35




\nhi Illingworth, G.~1977, ApJ, 218, L43



\nhi Jedrzejewski, R.~I.~1987, MNRAS, 226, 747

\nhi Kent, S.~M.~1990, in The Evolution of the Universe of Galaxies,
      ed.~R.~G.~Kron (San Francisco: ASP), 109





\nhi Kormendy, J.~1982, in Morphology and Dynamics of Galaxies, Twelfth 
      Advanced Course of the Swiss Society of Astronomy and Astrophysics,
      ed.~L.~Martinet and M.~Mayor (Sauverny: Geneva Obs.), 113




\nhi Kormendy, J.~1985, ApJ, 295, 73

\nhi Kormendy, J.~1987a, in IAU Symposium 127, Structure and Dynamics of
      Elliptical Galaxies, ed.~T.~de Zeeuw (Dordrecht: Reidel), 17

\nhi Kormendy, J.~1987b, in Nearly Normal Galaxies: From the Planck Time to 
      the Present, ed.~S.~M.~Faber (New York: Springer-Verlag), 163

\nhi Kormendy, J.~1988a, in Supermassive Black Holes, ed. M. Kafatos
      (Cambridge: Cambridge Univ.~Press), 98


\nhi Kormendy, J.~1988b, ApJ, 325, 128 

\nhi Kormendy, J.~1988c, ApJ, 335, 40 



\nhi Kormendy, J.~1992a, in Testing the AGN Paradigm, ed.~S.~S.~Holt, 
     S.~G.~Neff, \& C.~M.~Urry (New York: American Institute of Physics), 23

\nhi Kormendy, J.~1992b, in High Energy Neutrino Astrophysics, 
     ed.~V.~J.~Stenger, J.~G.~Learned, S.~Pakvasa, \& X.~Tata (Singapore:
     World Scientific), 196

\nhi Kormendy, J.~1993, in The Nearest Active Galaxies, ed.~J.~Beckman,
     L.~Colina \& H.~Netzer (Madrid: Consejo Superior de Investigaciones 
     Cient\'\i ficas), 197

\nhi Kormendy, J.~1994, in The Nuclei of Normal Galaxies: Lessons From The
     Galactic Center, ed.~R.~Genzel \& A.~I.~Harris (Dordrecht: Kluwer), 379

\nhi Kormendy, J., {\mit et al.}~1996a, ApJ, 459, L57  

\nhi Kormendy, J., {\mit et al.}~1996b, ApJ, 473, L91  

\nhi Kormendy, J., {\mit et al.}~1997, ApJ, 482, L139 

\nhi Kormendy, J., \& Bender, R.~1996, ApJ, 464, L119


\nhi Kormendy, J., \& Illingworth, G.~1982, ApJ, 256, 460 


\nhi Kormendy, J., \& Richstone, D.~1992, ApJ, 393, 559

\nhi Kormendy, J., \& Richstone, D.~1995, ARA\&A, 33, 581 (KR95)


\nhi Kormendy, J., \& Westpfahl, D.~J.~1989, ApJ, 338, 752

\nhi Lauer, T.~R.~1985, ApJS, 57, 473


\nhi Lauer, T.~R., {\mit et al.}~1995, AJ, 110, 2622


\nhi Lauer, T.~R., {\mit et al.}~1997, in preparation

\nhi Leach, R.~1981, ApJ, 248, 485






\nhi Michard, R., \& Simien, F.~1988, A\&AS, 74, 25

\nhi Miyoshi, M., {\mit et al.}~1995, Nature, 373, 127

\nhi Moffat, A.~F.~J.~1969, A\&A, 3, 455

\nhi Morgan, W.~W., \& Keenan, P.~C.~1973, ARA\&A, 11, 29

\nhi Nieto, J.-L., \& Bender, R.~1989, A\&A, 215, 266

\nhi Nieto, J.-L., Capaccioli, M., \& Held, E.~V.~1988, A\&A, 195, L1

\nhi Nieto, J.-L., Poulain, P., Davoust, E., \& Rosenblatt, P.~1991, A\&AS, 88,
      559

\nhi Peletier, R.~F., Davies, R.~L., Illingworth, G.~D., Davis, L.~E., \&
      Cawson, M.~1990, AJ, 100, 1091

\nhi Persson, S.~E., Frogel, J.~A., \& Aaronson, M.~1979, ApJS, 39, 61

\nhi Pierce, M.~1991, private communication

\nhi Poulain, P.~1988, A\&AS, 72, 215

\nhi Qian, E.~E., de Zeeuw, P.~T., van der Marel, R.~P., \& Hunter, C.~1995,
      MNRAS, 274, 602 







\nhi Richstone, D., {\mit et al.}~1998a, in preparation

\nhi Richstone, D., {\mit et al.}~1998b, in preparation



\nhi Salmon, D.~1985, CFHT Info.~Bull., No.~13, 9

\nhi Sandage, A.~1961, The Hubble Atlas of Galaxies (Washington, Carnegie
      Institution of Washington)

\nhi Sandage, A., Freeman, K.~C., \& Stokes, N.~R.~1970, ApJ, 160, 831

\nhi Sandage, A., \& Visvanathan, N.~1978, ApJ, 223, 707


\nhi Sargent, W.~L.~W., Schechter, P.~L., Boksenberg, A., \& Shortridge, 
      K.~1977, ApJ, 212, 326




\nhi Schechter, P.~L., \& Gunn, J.~E.~1979, ApJ, 229, 472


\nhi Scorza, C., \& Bender, R.~1995, A\&A, 293, 20









\nhi Stover, R.~J.~1988, in Instrumentation for Ground-Based Optical Astronomy:
     Present and Future, ed.~L.~B.~Robinson (New York: Springer-Verlag), 443

\nhi Strom, S.~E., Strom, K.~M., Goad, J.~W., Vrba, F.~J., \& Rice, W.~1976, 
      ApJ, 204, 684


\nhi Tody, D.,~{\mit et al.}~1986, IRAF User Handbook, Tucson, National Optical
     Astronomy Observatories

\nhi Tonry, J.~L.~1984, ApJ, 283, L27

\nhi Tonry, J.~L.~1987, ApJ, 322, 632

\nhi Tremblay, B., \& Merritt, D.~1995, AJ, 110, 1039

\nhi Tully, R.~B.~1988, Nearby Galaxies Catalog (Cambridge: Cambridge
      University Press)


\nhi van den Bosch, F.~C., \& Jaffe, W.~1997, in The Nature of Elliptical 
     Galaxies, ed.~M.~Arnaboldi, G.~S.~Da Costa, \&  P.~Saha (San Francisco:
     ASP), 142

\huge

\huge

\huge

\huge

\huge

\huge

\huge

\huge

\huge

\huge

\huge

\huge

\huge

\huge

\huge

\huge

\huge

\huge

\huge

\huge

\huge

\huge

\huge

\huge

\huge

\huge

\huge

\huge

\huge

\huge

\huge

\huge

\huge

\huge

\huge

\huge

\huge

\huge

\huge

\huge

\huge

\huge

\huge

\huge

\huge

\huge

\huge

\huge

\huge

\huge

\huge

\huge

\huge


\nhi van der Marel, R.~P., Cretton, N., de Zeeuw, P.~T., \& Rix, H.-W.~1997b,
      ApJ, submitted

\nhi van der Marel, R.~P., de Zeeuw, P.~T., Rix, H.-W.~1997, ApJ, in press

\nhi van der Marel, R.~P., de Zeeuw, P.~T., Rix, H.-W., \& Quinlan, G.~D.~1997a,
      Nature, 385, 610

\nhi van der Marel, R.~P., Evans, N.~W., Rix, H.-W., White, S.~D.~M., de
     Zeeuw, T.~1994b, MNRAS, 271, 99 

\nhi van der Marel, R.~P., \& Franx, M.~1993, ApJ, 407, 525

\nhi van der Marel, R.~P., Rix, H.-W., Carter, D., Franx, M., White, S.~D.~M.,
     de Zeeuw, T.~1994a, MNRAS, 268, 521 


\nhi Walker, G.~A.~H., {\mit et al.}~1984, in State-of-the-Art Imaging Arrays 
      and Their
      Applications, ed.~K.~N. Prettyjohns, Proc.~SPIE, 501, 353


\nhi Webb, C.~J.~1964, AJ, 69, 442


\nhi Whitmore, B.~C., McElroy, D.~B., \& Tonry, J.~L.~1985, ApJS, 59, 1


\end